\documentclass[10pt,fleqn,a4paper]{article}
\usepackage{cite}
\usepackage[numbers]{natbib}
\usepackage{mathptmx} 
\usepackage[T1]{fontenc}
\usepackage{parskip}
\usepackage[labelfont=bf,footnotesize]{caption}
\usepackage{amsmath,amssymb,amsfonts,amsthm}
\usepackage{graphicx}
\usepackage{tabularx}
\usepackage{setspace}                
\usepackage{wrapfig}
\usepackage{url}
\usepackage{lastpage}
\usepackage{multirow}
\usepackage{rotating}
\usepackage{adjustbox}
\usepackage{pdflscape}
\usepackage[svgnames,rgb,table]{xcolor}
\usepackage{tikz}
\usepackage[a4paper,hdivide={25mm,*,25mm},vdivide={25mm,*,25mm},headheight=14pt]{geometry}

\usepackage[textsize=small]{todonotes} 
\usepackage{tcolorbox}
\usepackage{titlesec}
\usepackage{ulem}
\usepackage{etoolbox}

\makeatletter
\patchcmd{\ttlh@hang}{\parindent\z@}{\parindent\z@\leavevmode}{}{}
\patchcmd{\ttlh@hang}{\noindent}{}{}{}
\makeatother

\usepackage[hidelinks]{hyperref}
\usepackage{fancyvrb}
\usepackage{lmodern}

\usepackage{fancyhdr}
\usepackage{array}
\newcolumntype{L}[1]{>{\raggedright\let\newline\\\arraybackslash\hspace{0pt}}p{#1}}
\newcolumntype{C}[1]{>{\centering\let\newline\\\arraybackslash\hspace{0pt}}p{#1}}
\newcolumntype{R}[1]{>{\raggedleft\let\newline\\\arraybackslash\hspace{0pt}}p{#1}}

\titleclass{\subsubsubsection}{straight}[\subsection]

\newcounter{subsubsubsection}[subsubsection]
\renewcommand\thesubsubsubsection{\thesubsubsection.\arabic{subsubsubsection}}

\titleformat{\subsubsubsection}
  {\normalfont\normalsize\bfseries}{\thesubsubsubsection}{1em}{}
\titlespacing*{\subsubsubsection}
{0pt}{3.25ex plus 1ex minus .2ex}{1.5ex plus .2ex}

\makeatletter

\def\toclevel@subsubsubsection{4}

\def\l@subsubsubsection{\@dottedtocline{4}{7em}{4em}}

\makeatother

\newenvironment{exampleoutput}[1]
 {\VerbatimEnvironment
  \begin{center}
  \tcbset{colback=blue!5!white,colframe=blue!75!black,title=\textbf{File contents example: \texttt{#1}}}
\begin{tcolorbox}
  \begin{minipage}{\linewidth}
  \begin{footnotesize}
  \begin{Verbatim}}
 {\end{Verbatim}
 \end{footnotesize}
  \end{minipage}
  \end{tcolorbox}
  \end{center}}

\newenvironment{bashinput}
 {\VerbatimEnvironment
  \begin{center}
  \tcbset{colback=orange!5!white,colframe=orange!75!white,title=\textbf{Bash input}}
\begin{tcolorbox}
  \begin{minipage}{\linewidth}
  \begin{footnotesize}
  \begin{Verbatim}}
 {\end{Verbatim}
 \end{footnotesize}
  \end{minipage}
  \end{tcolorbox}
  \end{center}}

\begin{document}
\parindent0mm

\pagestyle{plain}
\pagenumbering{arabic}

\begin{center}\Large
	{\bf Navigating Eukaryotic Genome Annotation Pipelines: A Route Map to BRAKER, Galba, and TSEBRA}\\
	\medskip
	\large
	Tomáš Brůna$^{1\ddag}$, Lars Gabriel$^{2,3}$, Katharina J. Hoff$^{2,3\ddag*}$\\
 \end{center}
 \small
 	$^{1}$) U.S.~Department of Energy Joint Genome Institute, Lawrence Berkeley National Laboratory, Berkeley, CA, 94720, USA\\
	$^{2}$) University of Greifswald, Institute of Mathematics and Computer Science, Walther-Rathenau-Stra\ss{}e 47, 17489 Greifswald, Germany\\
	$^{3}$) University of Greifswald, Center for Functional Genomics of Microbes, Felix-Hausdorff-Stra\ss{}e 8, 17489 Greifswald, Germany\\
	 $^{*}$) Senior author is also the first author.\\
    $^{\ddag}$) Corresponding authors, emails: tbruna@lbl.gov, katharina.hoff@uni-greifswald.de

\normalsize

\includegraphics[width=\linewidth]{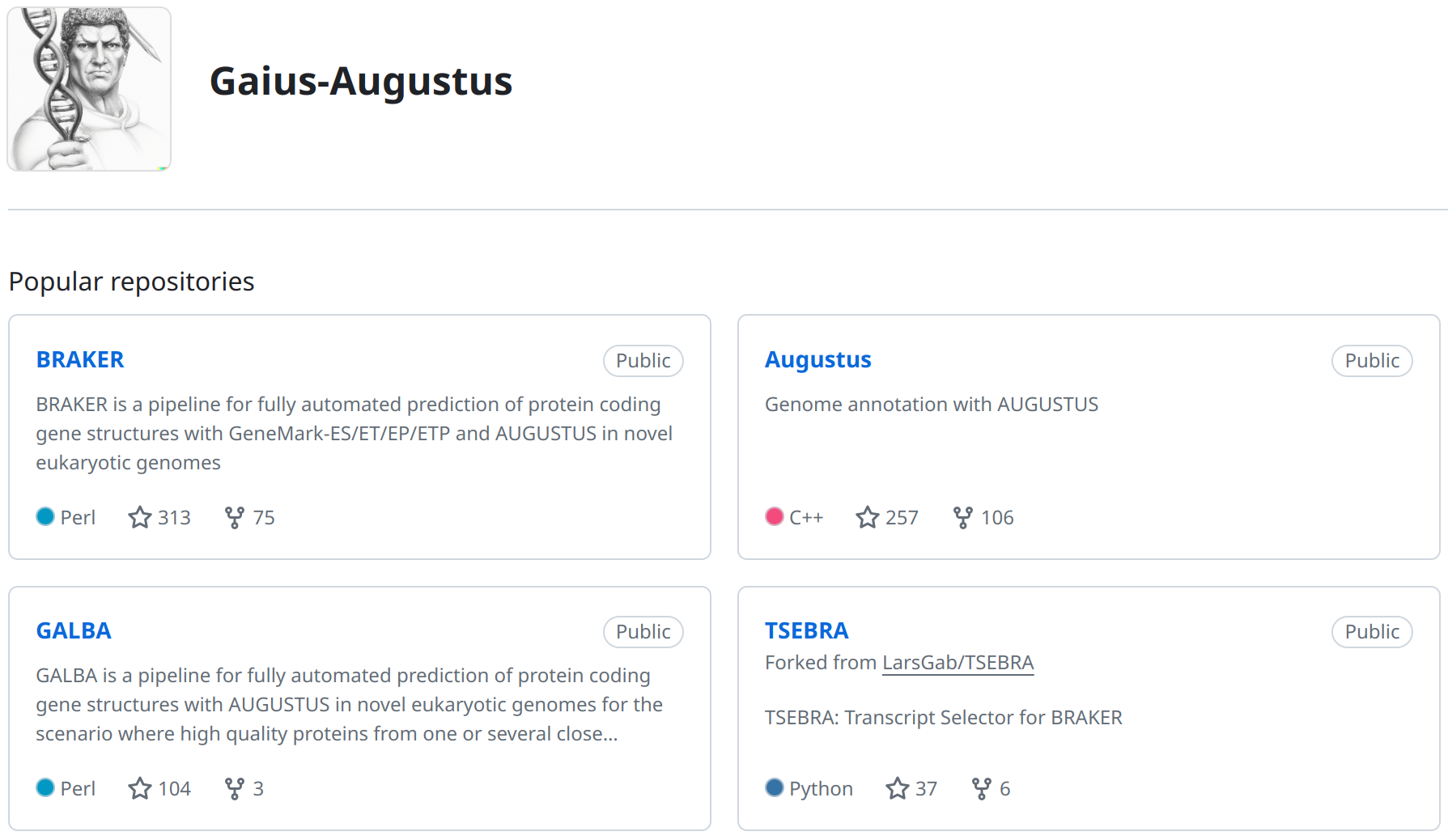}
\includegraphics[width=\linewidth]{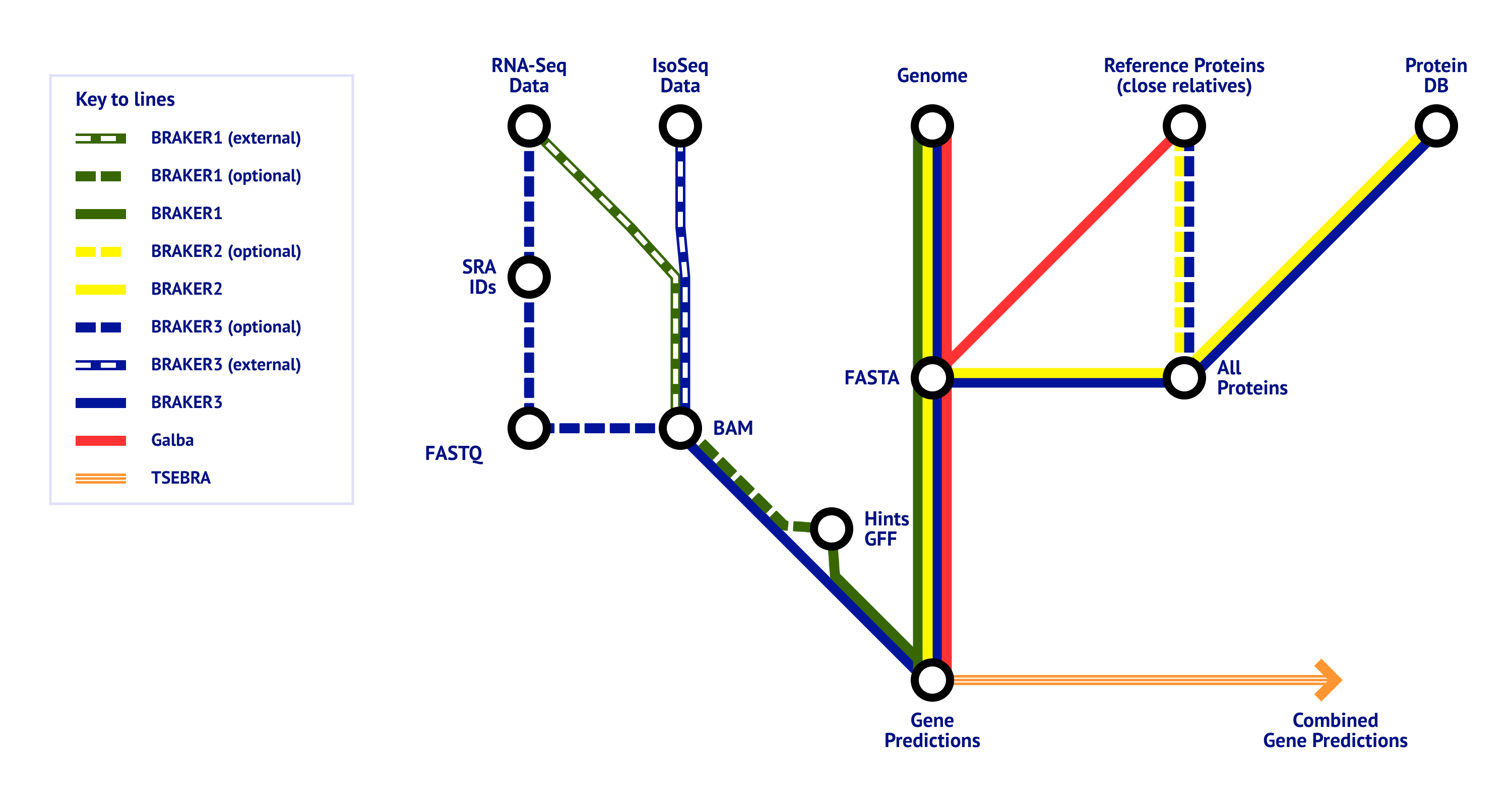}

\newpage
\tableofcontents
\newpage

\section*{Abstract}
Annotating the structure of protein-coding genes represents a major challenge in the analysis of eukaryotic genomes. This task sets the groundwork for subsequent genomic studies aimed at understanding the functions of individual genes. BRAKER and Galba are two fully automated and containerized pipelines designed to perform accurate genome annotation. BRAKER integrates the GeneMark-ETP and AUGUSTUS gene finders, employing the TSEBRA combiner to attain high sensitivity and precision. BRAKER is adept at handling genomes of any size, provided that it has access to both transcript expression sequencing data and an extensive protein database from the target clade. In particular, BRAKER demonstrates high accuracy even with only one type of these extrinsic evidence sources, although it should be noted that accuracy diminishes for larger genomes under such conditions. In contrast, Galba adopts a distinct methodology utilizing the outcomes of direct protein-to-genome spliced alignments using miniprot to generate training genes and evidence for gene prediction in AUGUSTUS. Galba has superior accuracy in large genomes if protein sequences are the only source of evidence. This chapter provides practical guidelines for employing both pipelines in the annotation of eukaryotic genomes, with a focus on insect genomes.

\section*{Keywords}
protein-coding genes, genome annotation, pipeline,
gene prediction,  AUGUSTUS, GeneMark-ETP, RNA-Seq, spliced alignment, miniprot, BRAKER, Galba, Iso-Seq, TSEBRA, BUSCO, compleasm

\section{Introduction}\label{introduction}


Genome annotation is the process of marking regions within an assembled genomic sequence with information about their structure and function. It identifies functional components, such as protein-coding genes, determines their structures, and classifies their functions. This fundamental step in genomic research supports the exploration of genes and proteins using molecular biology methods. 

\begin{figure}[h!]
	\begin{center}
		\begin{tabular}{lclc}
			A) & 
			\includegraphics[height=0.6\linewidth]{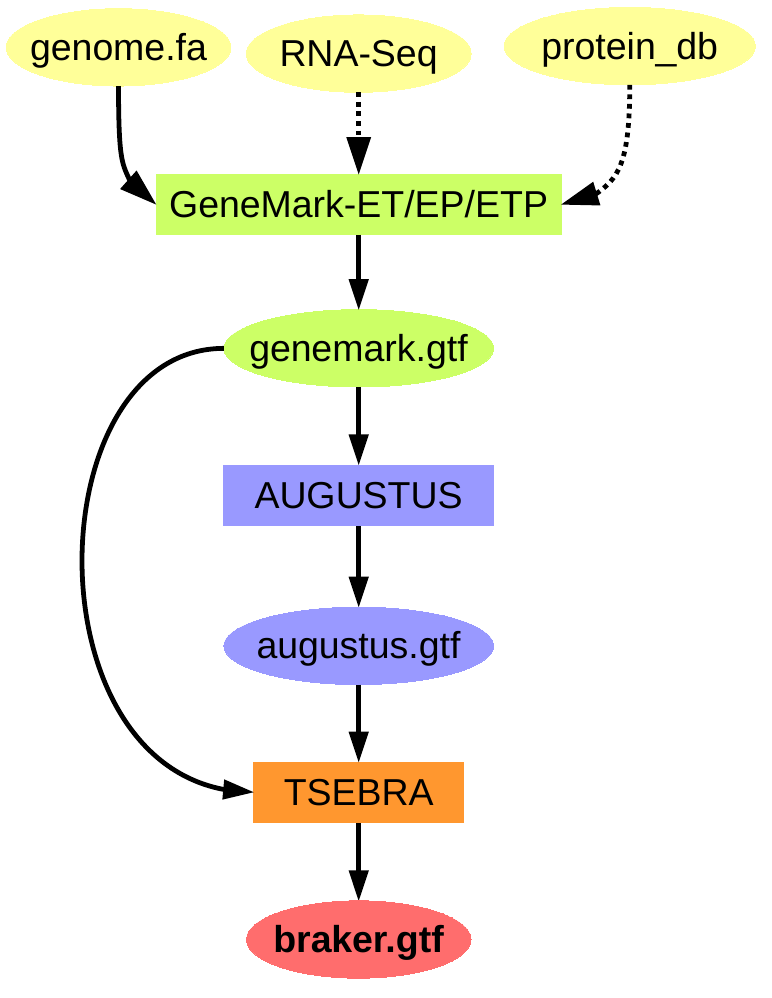} & B) & \includegraphics[height=0.6\linewidth]{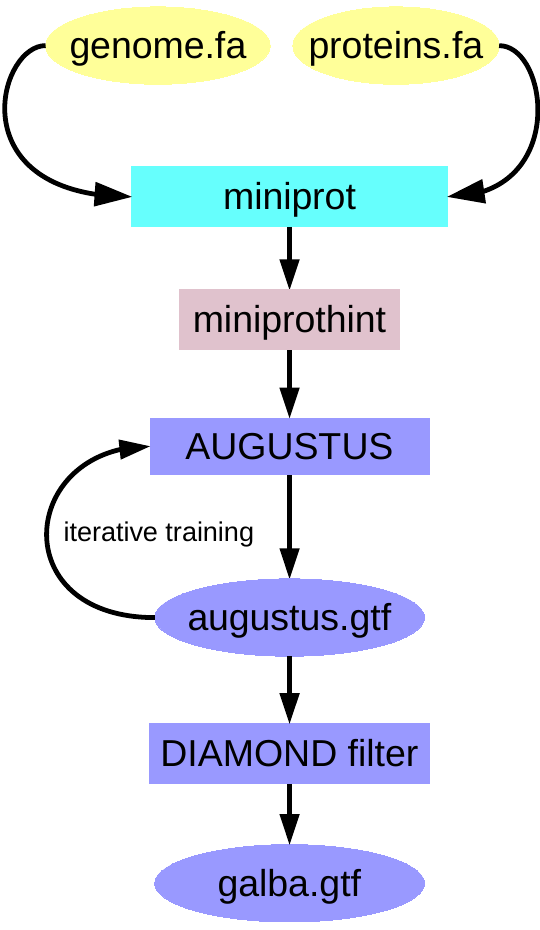}
		\end{tabular}
		
		\caption{Schematic view of the BRAKER \cite{braker15,bruuna2021braker2,gabriel2023braker3} and Galba \cite{bruuna2023galba} pipelines. A: In BRAKER, GeneMark-ET, -EP, or -ETP \cite{GeneMarkET2014,bruuna2020genemark,bruna2023genemark} is trained (using extrinsic data upon availability) and used to predict an initial set of genes (genemark.gtf). This set of genes is filtered, and the resulting high-quality genes are used to train AUGUSTUS \cite{MarioUCSC}. AUGUSTUS predictions are made with the derived species-specific parameters, using extrinsic data where available. Finally, TSEBRA \cite{gabriel2021tsebra} combines the set of AUGUSTUS genes with evidence-supported GeneMark genes to produce the \texttt{braker.gtf} output file. B: In Galba, miniprot \cite{li2023protein} is used to quickly splice align the protein sequences of a few closely related species to the genome. Miniprothint filters the alignments for quality and identifies training genes that are then used to train AUGUSTUS. AUGUSTUS predicts genes using filtered protein evidence. The initial predictions undergo quality filtering and are used for a second training iteration. In the last step, AUGUSTUS predicted genes are searched for DIAMOND \cite{buchfink2015fast} hits in the input protein data set to reduce false positive predictions.\label{braker-general}}
	\end{center}
\end{figure}

BRAKER \cite{braker15,bruuna2021braker2,gabriel2023braker3} and Galba \cite{bruuna2023galba} are fully automated and containerized bioinformatic pipelines for the prediction of protein-coding gene structures in eukaryotic genomes (see Figure \ref{braker-general}). Both BRAKER and Galba have undergone critical comparisons with competing pipelines, such as MAKER2 \cite{MAKER2} or FunAnnotate \cite{palmer2020funannotate}. In these experiments, BRAKER and Galba were shown to achieve superior accuracy.

\subsection{History of BRAKER and Galba}

The prototype of BRAKER was implemented in 2014 as a result of the international collaboration between the group of Mark Borodovsky at Georgia Tech Institute of Technology and the group of Mario Stanke at the University of Greifswald. Initially, BRAKER1 \cite{braker15} connected the GeneMark-ET self-training gene finder \cite{GeneMarkET2014} with the AUGUSTUS gene finder \cite{MarioUCSC} that requires supervised training on gene examples of a target species. GeneMark-ET uses RNA-Seq read spliced alignments for its training step and is used to generate a set of training genes for AUGUSTUS, which can use RNA-Seq information during its gene prediction step. The final output of BRAKER1 was originally an AUGUSTUS gene set. The accuracy of BRAKER1 clearly exceeded the state-of-the-art when the pipeline was first published in 2016. 

Supported by a National Institutes of Health grant to Mark Borodovsky and Mario Stanke, the pipeline was subsequently extended to connect the GeneMark-EP/EP+ self-training gene finder \cite{bruuna2020genemark} to AUGUSTUS, resulting in BRAKER2 \cite{bruuna2021braker2}. BRAKER2 uses a large database of clade-specific proteins to achieve high precision, even if proteins from close relatives are lacking from the protein database. In particular, BRAKER2 does not simultaneously process RNA-Seq and protein information.

Both BRAKER1 and BRAKER2 suffer from limited accuracy in large genomes (>1 Gbp) where they tend to predict fragmented genes.

With the availability of BRAKER1 to take advantage of RNA-Seq data and BRAKER2 to utilize a large database of clade-specific proteins, TSEBRA, the transcript selector for BRAKER, was developed to optimally combine their predictions. However, as the combination of RNA-Seq and protein inputs is now better solved by BRAKER3 (see below), the use of TSEBRA has evolved to mainly combine prediction sets within BRAKER runs. Furthermore, at the time of the development of TSEBRA, BRAKER also supported several alternative pipelines centered on the GenomeThreader protein-to-genome alignment tool \cite{gremme2013computational}. These alternative pipelines, described in \cite{hoff2019whole}, have been discontinued because their accuracy was not competitive after BRAKER2's release. However, one of the GenomeThreader-based pipelines later gave rise to Galba (see below).

In 2023, BRAKER3 became available \cite{gabriel2023braker3}. BRAKER3 is the most accurate and yet most flexible of the BRAKER pipelines. Combining the GeneMark-ETP gene finder \cite{bruna2023genemark} with AUGUSTUS, BRAKER3 can simultaneously process RNA-Seq data and information from a large protein database, achieving greater accuracy than previous pipelines in genomes of various sizes. (Please note that the BRAKER developers have never tested a run on extremely large genomes, such as the mistletoe; large refers to a few Gbp in this context.) BRAKER3 uses TSEBRA to combine the output of GeneMark-ETP and AUGUSTUS predictions into an optimal gene set. With the release of BRAKER3, BRAKER1, and BRAKER2 were also modified to internally use TSEBRA to produce a combined set of genes from AUGUSTUS and GeneMark predictions (see Figure \ref{braker-general}).

When developing BRAKER, we primarily used well-annotated model genomes to allow accurate evaluation with respect to a high-quality reference annotation. In BRAKER2 and BRAKER3, we added more genomes with possibly less reliable annotations for comparison purposes. Still, in real-life applications of BRAKER3 where sufficient extrinsic data may be lacking, we and also independent users quickly found that the TSEBRA combiner using its default BRAKER parameters may be too conservative; i.e.,~it may discard correct transcripts with BUSCO \cite{seppey2019busco} support. BUSCO, short for Benchmarking Universal Single-Copy Orthologs, serves as a tool to assess genomes, proteins, and transcripts for the presence of these single-copy orthologs and as a term denoting sets of clade-specific marker genes. In extreme scenarios, BRAKER3 may simply discard a large fraction of correctly predicted transcripts. In January 2024, we presented a possible solution to this problem at the Plant and Animal Genome Conference in San Diego. BRAKER was extended to assess BUSCO completeness of the underlying gene set with a compleasm \cite{huang2023compleasm} version that was extended by the lead author Neng Huang for the purpose of executing HMMER \cite{mistry2013challenges} on a protein sequence set. We implemented a decision scheme (see Figure \ref{compleasm}) to force TSEBRA to retain transcripts that were identified as similar to BUSCOs by compleasm or even to enforce complete preservation of the best set of underlying transcripts if BUSCOs would otherwise drop too drastically. Typically, BUSCO serves as a sensitivity measure for final gene sets against clade-specific marker genes. However, here we use the BUSCO data sets to ensure that such marker genes end up in the final predictions, because that is ultimately of interest for possible downstream experiments. 

As mentioned above, BRAKER historically included several pipelines centered on the GenomeThreader alignment tool rather than the GeneMark gene finder. The core functionality of the purely protein-centered GenomeThreader pipeline was transferred to the new Galba spin-off pipeline \cite{bruuna2023galba} in 2023.  Genome\-Threader pipelines, in contrast to GeneMark-centered BRAKER pipelines, always required the input of protein sequences from species that were closely related to achieve reasonable results. With the release of miniprot, a faster and more robust protein-to-genome spliced aligner \cite{li2023protein}, we replaced GenomeThreader with miniprot in Galba to take advantage of its superior speed and better accuracy with greater divergence between query proteins and the target genome. Galba underwent several further improvements since it branched off from BRAKER. In particular, we adopted protein evidence scoring and filtering described first for GeneMark-EP/EP+ and integrated Pygustus for parallelization during gene prediction with AUGUSTUS. Unlike BRAKER2, Galba is a pipeline that should run with a selected set of proteins from several close relatives. In principle, Galba can also run with a large database of clade-specific proteins, similar to BRAKER2, but this causes increased run-time, and in such scenarios, BRAKER2 is often more accurate. Galba generally outperforms BRAKER2 in large genomes (> 1Gbp). Galba cannot currently handle any transcriptome evidence; however, the resulting gene sets can easily be combined with e.g.~BRAKER1 gene sets with TSEBRA.

In this book chapter, we provide detailed hands-on instructions on the application of BRAKER, Galba, and TSEBRA to annotate protein-coding gene structures in eukaryotic genomes. Here we build on the previously published book chapter \textit{Whole Genome Annotation with BRAKER} \cite{hoff2019whole} and replace it. In part, we re-use e.g.~command lines if no updates are necessary.

As per user requests, we also document an interim solution for using high-quality Iso-Seq data. 

The volume encompassing this book chapter has a focus on insects. Although none of the tools discussed here are insect-specific pipelines, they are widely used to annotate insect genomes. Therefore, we provide a test data set from \textit{Drosophila melanogaster} for novices to practice using these tools. 

\subsection{Navigating this chapter}

This chapter of the book provides hands-on instructions. We use two types of boxes to illustrate file formats and commands. The blue boxes are used to illustrate the input and output file formats and contents.

\begin{exampleoutput}{File format}
This type of box is used to show you either file format or file content examples.
\end{exampleoutput}

The orange boxes show commands that can be executed in a Bash terminal, such as calling the BRAKER pipeline. The following example shows how to print \textit{ Hello world} in Bash with \texttt{echo}:

\begin{bashinput}
echo "Hello world"
\end{bashinput}

\section{Software and input files}

In this section, we describe the required computational resources, software, and input files for executing BRAKER, Galba, and TSEBRA. We provide two types of data sets for testing the pipelines: (1) small data sets that always come with the software to test whether an installation was successful, and (2) larger files from \textit{Drosophila} melanogaster to practice the pipelines. For some usage examples, we only provide the larger files (e.g.~it does not make sense to download only the first couple of reads that correspond to a Sequence Read Archive identifier because these particular reads may lead to a crash of the pipeline due to missing information).

\subsection{Computational resources}

Three types of resources should be considered for running BRAKER and Galba (TSEBRA requires very little resources): (1) CPU threads, (2) RAM, and (3) file system input/output (I/O) capacity. Regarding the minimal system requirements, BRAKER can in many cases be run effectively on a modern desktop computer equipped with 8 GB of RAM per thread, each container images require ~3 GB of free hard drive space (in addition to space for results). In the following, we will describe in more detail the factors considered to select a suitable number of threads, and the impact of pipeline on the file system.

\subsubsection{Threads}

BRAKER and Galba are parallelized pipelines. The number of threads can be adjusted by the user. The more threads are used, the faster the pipeline will run. However, BRAKER is not a pipeline that scales well on hundreds of threads.

There are two main phases in BRAKER and Galba: (1) training and (2) gene prediction, and both phases are data-parallelized (see section \ref{filesystem} below).

In BRAKER and Galba, AUGUSTUS training involves parameter optimization through $k$-fold cross-validation (default $k=8$). Training genes are repeatedly divided into $k$ parts during this procedure. BRAKER and Galba will ensure that each bucket has a minimum of 200 training genes. Excess threads will not be used in the training phase, they will be used in the gene prediction phase.

\subsubsection{File system \label{filesystem}}

The data parallelization approaches in both BRAKER and Galba can occasionally impact the speed of a file system when numerous pipeline jobs are executed concurrently on a High-Performance Compute system sharing a single file system. This aspect deserves consideration to prevent potential slowdowns where large-scale parallel job execution is common.

During the gene prediction phase, BRAKER and Galba employ different parallelization strategies. BRA\-KER splits the genome file into fragments corresponding to the number of contigs or scaffolds, which might overload the file system with too many files if the genome is highly fragmented. Therefore, BRAKER will issue a warning if AUGUSTUS predictions for more than 30,000 (sub-)sequences\footnote{AUGUSTUS is applied in batches to large sequences, producing a large number of prediction files prior to joining these predictions.} are to be expected. In such cases, it is safer to execute BRAKER with one thread. In contrast, Galba employs Pygustus for parallelization, dividing the genome file based on the number of threads and operating within the system's \texttt{/tmp/} directory to avoid file system overload, which is especially effective in high-performance computing (HPC) environments with dedicated \texttt{/tmp/} directories on individual nodes.

\subsection{Software}

BRAKER is available on GitHub at \url{https://github.com/Gaius-Augustus/BRAKER}, and Galba is available at \url{https://github.com/Gaius-Augustus/Galba}. The TSEBRA combiner is available at \url{https://github.com/Gaius-Augustus/TSEBRA}. All three tools are fully open source. Detailed instructions for manually setting up the complex environments of BRAKER and Galba are provided on their respective GitHub pages. Both pipelines depend on a large number of tools and libraries, requiring specific versions for some. To simplify installation, we have containerized both pipelines in Docker containers available on Dockerhub. The BRAKER container is available at \url{https://hub.docker.com/r/teambraker/braker3}, while the Galba container is available at \url{https://hub.docker.com/r/katharinahoff/galba-notebook/}. TSEBRA is included in both containers. 

The Docker containers can be executed with Docker itself. However, Docker requires root privileges, which may not be available to users on high-performance compute systems. The containers have therefore also been tested with Singularity, which is more commonly available on HPC systems and does not require root access. This book chapter assumes that users have access to a Linux system with Singularity.

Installing Singularity requires root permissions. We refer the reader to the excellent tutorial at \url{https://singularity-tutorial.github.io/01-installation/} for instructions. Users intending to work on an HPC system should consult their system administrator for Singularity installation.

At the time of writing, the current version of BRAKER is 3.0.8 whereas the current version of Galba is v1.0.11. TSEBRA is in version 1.1.2.3. 

\subsubsection{Pulling containers with Docker}

If you wish to execute BRAKER with Docker, pull the Docker container with:

\begin{bashinput}
sudo docker pull teambraker/braker3:latest
\end{bashinput}

The resulting container image is at the time of writing 2.97 GB in size. 

Similarly, if you want to execute Galba with Docker, pull the container as follows:

\begin{bashinput}
sudo docker pull katharinahoff/galba:latest
\end{bashinput}

At the time of writing, the image had 2.39 GB. We tested BRAKER and Galba with Docker version 24.0.4.

As of writing this book chapter, the production version of GeneMark-ETP does not support long-read transcriptome data. However, a modified GeneMark-ETP version that performs StringTie2 assembly on Iso-Seq data has been made available on GitHub and has been tested. To pull the BRAKER container with this GeneMark-ETP version that was designed for Iso-Seq transcriptome data, execute the following:

\begin{bashinput}
sudo docker pull teambraker/braker3:isoseq
\end{bashinput}

The image size is identical to the size of the standard BRAKER container. The only difference lies in the command line option to execute StringTie within GeneMark-ETP. It should be noted that this modified version of GeneMark-ETP, and thus this version of the BRAKER container, does not correctly process short read transcriptome data. 

\subsubsection{Pulling containers with Singularity}

For execution with Singularity, you need to build so-called Singularity image files from the Docker containers. To build such an image for BRAKER, execute the following:

\begin{bashinput}
singularity build braker3.sif docker://teambraker/braker3:latest
\end{bashinput}

The resulting file, \texttt{braker3.sif}, currently has a size of 2.3 GB.

To build a Singularity image file for Galba, execute:

\begin{bashinput}
singularity build galba.sif docker://katharinahoff/galba:latest
\end{bashinput}

The \texttt{galba.sif} file currently has a size of 2.0 GB. 

To build a Singularity image file for the BRAKER container for Iso-Seq data, execute the following:

\begin{bashinput}
singularity build braker3-isoseq.sif docker://teambraker/braker3:isoseq
\end{bashinput}

\subsection{Input files}

\begin{figure}[h!]
	\begin{center}
		\includegraphics[width=\linewidth]{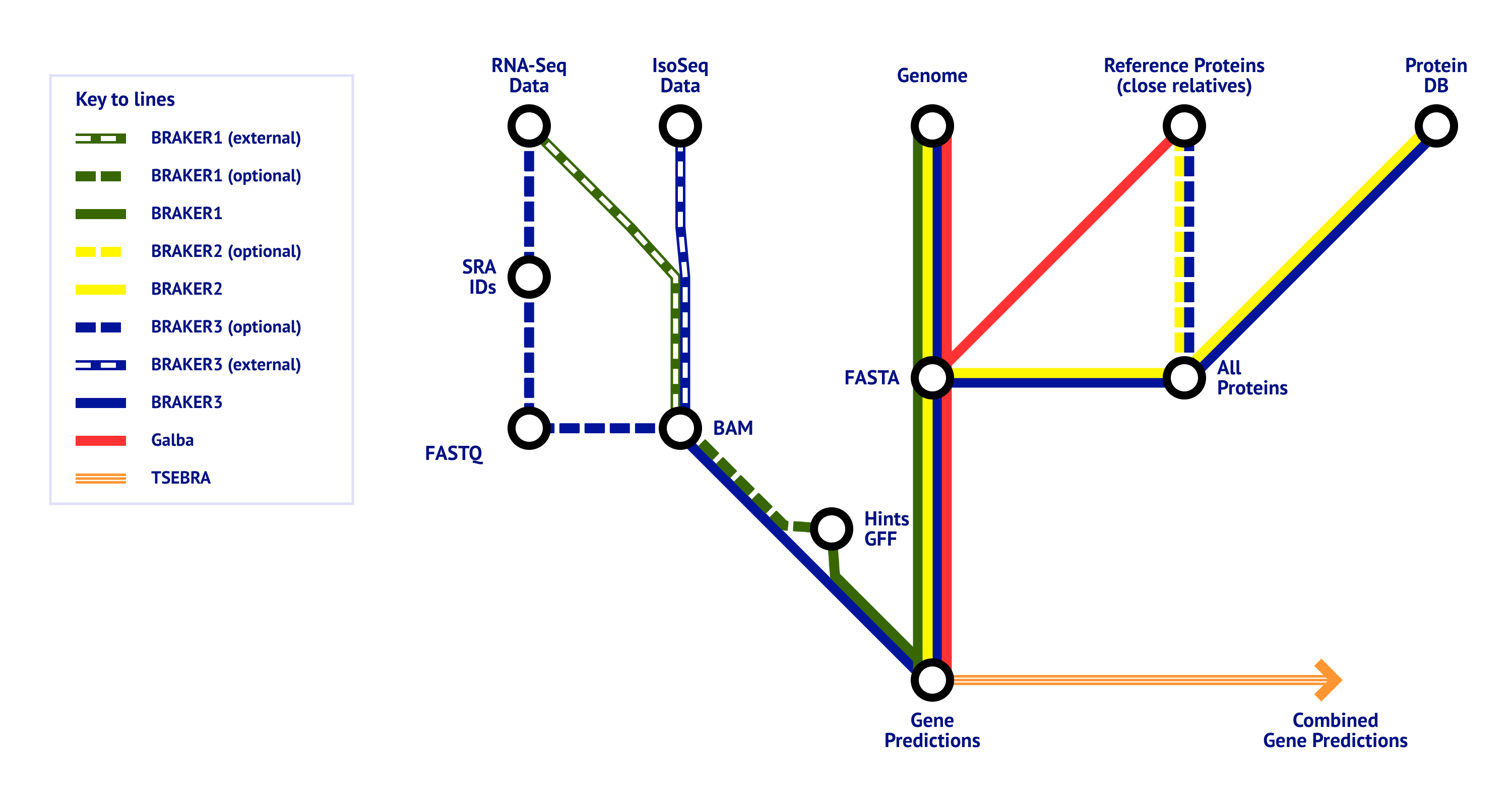}
		\caption{Input types and formats for BRAKER and Galba pipelines. The Genome input to all pipelines is always in FASTA format. For BRAKER1, users must prepare a BAM file with splice-aligned RNA-Seq data or an AUGUSTUS hints file in GFF format. BRAKER3 offers more input options for transcriptome: users may provide sequence read archive identifiers for libraries, a directory with FASTQ files, or the prepared BAM file, with the latter being the only possible input for BRAKER3 with Iso-Seq. The protein input for BRAKER2, BRAKER3, and Galba is in FASTA format. Note that BRAKER2 and BRAKER3 are commonly executed with a very large database of protein sequences, e.g.,~OrthoDB partitions. It is possible to augment these databases with full-genome annotation-scale protein sequences from close relatives. For Galba, only such reference proteins from close relatives should be used as input. The output of all BRAKER and Galba pipelines can be combined using the TSEBRA combiner for an improved set of genes. \label{subway}}
	\end{center}
\end{figure}

BRAKER and Galba are both pipelines for annotating eukaryotic genomes. Figure \ref{subway} 
illustrates the different input options for BRAKER1, BRAKER2, BRAKER3, and Galba, including their
respective file formats. The figure introduces a color scheme similar to a subway map, with the line's color scheme used throughout the entire book chapter to aid orientation.

In summary, both the BRAKER pipelines and Galba always require a genome
file in FASTA format. Galba, BRAKER2, and BRAKER3 require a protein sequence file in FASTA format, albeit with differing requirements for the protein file's content. BRAKER1 and BRAKER3 require transcriptome evidence as input and offer different possible starting points. In the following, we describe the different input options and, where applicable, their respective formats in more detail.

\subsubsection{Genome input for all pipelines \protect\includegraphics[width=0.8cm]{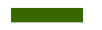}\protect\includegraphics[width=0.8cm]{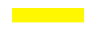}\protect\includegraphics[width=0.8cm]{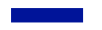}\protect\includegraphics[width=0.8cm]{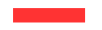}}

A genome file in FASTA format is a mandatory input for both BRAKER and Galba. For optimal results, 
this file should encompass the entire genome assembly. Only short contigs, which often contain mostly incomplete gene structures, are less suitable and may be removed. Although AUGUSTUS has the capability to predict partial genes within short sequences, its training process is based exclusively on full gene structures. Consequently, including short contigs in a Galba or BRAKER execution may not enhance the training of parameters and is likely to extend the overall runtime due to the necessity of processing each contig during all stages. To prevent extended run-time, it is advisable to omit contigs shorter than 3,000 nucleotides.

Below, you find an example of a miniature genome file in multiple FASTA format:

\begin{exampleoutput}{genome.fa}
>X1
CTGGTGACTATGAGGACATGCTTGTTGCTCTTCTCGGACATGGCGATGCTTGAAACTGTT
TCAACTTTCGAGTTCCTCCTTTCTCTTACTGCATGGTTTGTTTTAAATAAAAGAGTTGTG
AAACTGGTTCTGCAACTATTTATCAATGATCGTTTGAGTTTGTTAAATTTGAATC
>X2
AAGTGACGTGTTCGATGTACTATGGGTCCTACTCACATGGCTCTCCACCAATGTCTTTGT
CTGAAAGCGACACCGCAAAAGTCTTCTGGGCTCTACTTTGGGCCCTACCACAGCTTTTTG
GGTTTTTGGACTTTCTCTAAATCCAAAAGGTTCGATTAGAAATAGATCTCGTGCAAATCT
AACACGTGTACGGCAGCTACGTCTGGAGATCCATATGATCAGGATACTGATCGCCGCATG
TAATACGACCTCTCAC
\end{exampleoutput}

Please note that the FASTA headers shall not contain special characters or white spaces and must be unique across the genome file. The genome file should not contain any non-ATGCN characters and should be encoded with utf-8.

Eukaryotic genomes often feature repetitive elements, which can complicate gene prediction efforts as they frequently contain open reading frames similar to genuine coding genes. Consequently, misidentifying transposable elements as coding genes can result in inaccurate predictions and inflated gene counts. Therefore, it is highly recommended to thoroughly mask these repetitive sequences in the genome before initiating a Galba or BRAKER analysis. 

We recommend RepeatModeler2 \cite{flynn2020repeatmodeler2} and RepeatMasker \cite{UsingRepeatMasker} to generate a species-specific repeat library and mask the genome as illustrated in the following example (the required software is not included the BRAKER and Galba containers; we refer users to the repository \url{https://github.com/Dfam-consortium/TETools}):

\begin{bashinput}
T=72 # specify number of threads
GENOME=genome.fa # specify genome file
DB=some_db_name_that_fits_to_species # make up a name for the database

BuildDatabase -name ${DB} ${GENOME}
RepeatModeler -database ${DB} -pa ${T} -LTRStruct
RepeatMasker -pa ${T} -lib ${DB}-families.fa -xsmall ${GENOME}
\end{bashinput}

This results in a file \texttt{\$\{GENOME\}.masked} that can serve as input to BRAKER or Galba.

For large genomes, e.g.~vertebrates, the execution of RepeatModeler2 and RepeatMasker as shown above may result in a severe undermasking of repeats, as these tools are prone to missing tandem repeats with large repeat periods. We recommend an additional tandem repeat finder \cite{benson1999tandem} run with a larger maximum repeat period size (500) to achieve more complex masking. The following example shows how to run this (including data parallelization):

\begin{bashinput}
# The script splitMfasta.pl is from https://github.com/Gaius-Augustus/Augustus
splitMfasta.pl --minsize=25000000 ${GENOME}.masked

# Running TRF
ls genome.split.*.fa | parallel 'trf {} 2 7 7 80 10 50 500 -d -m -h'

# Parsing TRF output
# The script parseTrfOutput.py is from https://github.com/gatech-genemark/BRAKER2-exp
ls genome.split.*.fa.2.7.7.80.10.50.500.dat | parallel 'parseTrfOutput.py {} --minCopies 1 \ 
	--statistics {}.STATS > {}.raw.gff 2> {}.parsedLog'

# Sorting parsed output..."
ls genome.split.*.fa.2.7.7.80.10.50.500.dat.raw.gff | parallel 'sort -k1,1 -k4,4n -k5,5n {} \ 
	> {}.sorted 2> {}.sortLog'

# Merging gff...
FILES=genome.split.*.fa.2.7.7.80.10.50.500.dat.raw.gff.sorted
for f in $FILES
do
	bedtools merge -i $f | awk ' \ 
		BEGIN{OFS="\t"} {print $1,"trf","repeat",$2+1,$3,".",".",".","."}' \ 
		> $f.merged.gff 2> $f.bedtools_merge.log
done
	
# Masking FASTA chunk
ls genome.split.*.fa | parallel 'bedtools maskfasta -fi {} \ 
	-bed {}.2.7.7.80.10.50.500.dat.raw.gff.sorted.merged.gff -fo {}.combined.masked -soft \ 
	&> {}.bedools_mask.log'
	
# Concatenate split genome
cat genome.split.*.fa.combined.masked > genome.fa.combined.masked
\end{bashinput}

The resulting file \texttt{genome.fa.combined.masked} can serve as input to BRAKER or Galba.

RepeatModeler2 and RepeatMasker are time consuming. If users of BRAKER and Galba are not interested in repeats themselves, RED \cite{girgis2015red} can be a much faster alternative. Here is how to run RED (software not contained in the BRAKER and Galba containers, find installation instructions at \url{https://github.com/nextgenusfs/redmask}):

\begin{bashinput}
redmask.py -i genome.fa -o genome
\end{bashinput}

For managing repeat sequences within a genome, two approaches are commonly employed: \textit{soft-masking} and \textit{hard-masking}. Soft masking involves denoting the repeat regions of a genome in a FASTA file with lowercase letters, whereas uppercase letters indicate nonrepeat sequences. On the contrary, hard-masking replaces every nucleotide identified as part of a repeat with the letter \texttt{N}. Hard-masked nucleotides remain permanently invisible to gene finders employed in BRAKER and Galba; i.e., any overlaps of repeats with protein-coding genes will prevent the genes from being correctly predicted.
 
GeneMark-ETP, the underlying tool of BRAKER, and AUGUSTUS, the underlying tool of both BRAKER and Galba, have been carefully developed to utilize soft-masked repeats. Thus,  we recommend that repeats are \textit{soft-masked} for running BRAKER and Galba.

For BRAKER and Galba, alignment files that link evidence to the genome sequence play a crucial role. These files must reference the names of genomic sequences exactly as they appear in the FASTA headers of the genome file. However, complications arise when alignment tools encounter long or complex FASTA headers, often leading to truncated sequence names. This discrepancy between the sequence names in the alignment files and those in the genome file can impede BRAKER's functionality, which necessitates identical naming. To avoid such issues, it is advisable to examine and simplify the FASTA headers in your genome file before using any alignment tools or running BRAKER. For simplification, scripts like the one available at \url{http://bioinf.uni-greifswald.de/bioinf/downloads/simplifyFastaHeaders.pl} can be used to generate a new genome file with simplified headers, alongside a mapping table to cross-reference the original and updated headers.

\begin{bashinput}
simplifyFastaHeaders.pl in.fa prefix out.fa header.map
\end{bashinput}

\subsubsection{Transcriptome input for BRAKER1 and BRAKER3}

BRAKER1 and BRAKER2 accept short read RNA-Seq evidence in BAM format. BRAKER1 alternatively accepts a GFF-formatted hints file. BRAKER3 has several alternative upstream starting points: Sequence Read Archive \cite{katz2022sequence} identifiers (SRA IDs) or a directory with FASTQ (or BAM) files.

BRAKER3 can also process PacBio HiFi Iso-Seq data if a BAM file is prepared beforehand. However, a
custom modified version of GeneMark-ETP is required to process the BAM file. This is not included
in the standard BRAKER3 container (see section \ref{isoseq}), and it is not possible to mix short and long read transcriptome evidence in a single BRAKER3 run.

\subsubsubsection{Short read BAM files \protect\includegraphics[width=0.8cm]{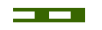}\protect\includegraphics[width=0.8cm]{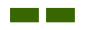}\protect\includegraphics[width=0.8cm]{figs/braker3.png}}

\textbf{BRAKER1} and \textbf{BRAKER3} accept alignment files of RNA-Seq reads mapped against the target genome in BAM format. Since BRAKER uses the information of how many reads cover a particular potential splice site in the genome, it is crucial that only RNA-Seq data producing high read coverage are used, and that only aligners producing spliced alignments are employed. RNA-Seq data produced by sequencing methods that generate a high number of (short) reads from mRNAs (e.g.~Illumina) are therefore particularly useful for running BRAKER1. We recommend using HiSat2 \cite{kim2019graph} for a short read-to-genome spliced alignment. The following example shows how to align RNA-Seq reads to a genome with HiSat2:

\begin{bashinput}
T=72 # specify number of threads
hisat2-build genome.fa genome
hisat2 -p ${T} -q -x genome -1 reads_1.fq -2 reads_2.fq -S rnaseq.sam
\end{bashinput}

This SAM format file must be converted to the BAM format with Samtools \cite{danecek2021twelve}:

\begin{bashinput}
samtools view -Sb rnaseq.sam > rnaseq.bam
\end{bashinput}

RNA-Seq data are not necessarily generated to improve gene prediction accuracy. Large volumes of RNA-Seq are used in gene expression studies, often involving several biological and/or technical replicates for statistical purposes. It usually does not do any harm to use all available RNA-Seq data, but it increases alignment and BRAKER runtime and memory requirements. You may obtain highly similar results if you restrict yourself to using e.g.~one randomly selected replicate of each library type.

Sometimes, you may wish to use RNA-Seq data that were deposited in the Sequence Read Archive by third parties. If there are large amounts of RNA-Seq libraries for your species, we recommend VARUS \cite{stanke2019varus} to automatically sample and align the RNA-Seq data.

\subsubsubsection{Long read BAM files \protect\includegraphics[width=0.8cm]{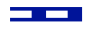}\protect\includegraphics[width=0.8cm]{figs/braker3.png}}\label{isoseq}

\textbf{BRAKER3} can process PacBio HiFi Iso-Seq data if a BAM file is prepared beforehand. However, this requires a custom modified version of GeneMark-ETP, only included in a separately tagged container (see \ref{isoseq}). The following example shows how to align Iso-Seq data to a genome with minimap2 \cite{li2018minimap2}:

\begin{bashinput}
T=72 # specify number of threads
minimap2 -t ${T} -ax splice:hq genome.fa isoseq.fa > isoseq.sam
samtools view --threads ${T} -Sb isoseq.sam > isoseq.bam
\end{bashinput}

The resulting file \texttt{isoseq.bam} can be used as input to BRAKER3 if the Iso-Seq container is used.

\subsubsubsection{Hints file in GFF format \protect\includegraphics[width=0.8cm]{figs/braker1.png}}

Instead of providing RNA-Seq alignments in BAM format to \textbf{BRAKER1}, users can provide a hints file generated by \texttt{bam2hints} from the BAM file. This file, which contains information extracted from the BAM file, is the initial output produced by \texttt{braker.pl} when a BAM file is given on input. There are two main reasons why users may want to execute \texttt{bam2hints} to generate hints from BAM files before initiating BRAKER:

\begin{enumerate}
	\item Efficiency through parallel processing: Although BRAKER can process \texttt{bam2hints} across multiple BAM files concurrently, it does so using only the number of threads allocated to the entire BRAKER process. When dealing with a large number of BAM files and having additional threads that should not be dedicated to BRAKER, executing \texttt{bam2hints} on these additional threads can lead to a decrease in overall processing time.
	\item Reduction in file size: The hint files produced are significantly smaller than the original BAM files. In certain computing environments, particularly those utilizing virtual resources, minimizing file size may be beneficial for performance and storage reasons.
\end{enumerate}

Hints files are formatted in a nine-column tab-separated GFF structure. An example of this format is produced by the AUGUSTUS utility \texttt{bam2hints}, utilized by BRAKER to create hints from RNA-Seq BAM files. In the following, we show you how to sort a BAM file, how to run \texttt{bam2hints} on the sorted file, and how the output format looks like:

\begin{bashinput}
T=72 # specify number of threads
singularity exec -B ${PWD}:${PWD} braker3.sif samtools sort \
  -@${T} -o rnaseq.s.bam rnaseq.bam
singularity exec -B ${PWD}:${PWD} braker3.sif bam2hints \
  --intronsonly --in=rnaseq.s.bam --out=hints.gff
\end{bashinput}

\begin{exampleoutput}{RNAseq.hints}
2R b2h intron 336478 343473 0 . . mult=3;pri=4;src=E
2R b2h intron 336480 343473 0 . . mult=11;pri=4;src=E
2R b2h intron 336482 343473 0 . . mult=2;pri=4;src=E
2R b2h intron 336658 427382 0 . . pri=4;src=E
\end{exampleoutput}

Note that the last column contains a field \texttt{mult=INT}. This indicates the coverage information for a feature, e.g.~the given intron in line 1 has support from 3 RNA-Seq reads. The field \texttt{src=E} indicates that this feature was extracted from the expression data. The source tags correspond to an AUGUSTUS configuration file that contains weights on how to treat evidence from this particular source.

The third column of the hints file contains the feature name of a hint. In BRAKER1, only \texttt{intron} hints from RNA-Seq are supported. The features \texttt{ass} and \texttt{dss} (donor and acceptor splice sites) are automatically derived from \texttt{intron} hints by AUGUSTUS.

It is important to note that \textbf{BRAKER3} processes RNA-Seq evidence differently, providing the hints file produced by \texttt{bam2hints} only works for \textbf{BRAKER1}.

\subsubsubsection{Sequence Read Archive identifiers \protect\includegraphics[width=0.8cm]{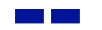}}

If the RNA-Seq data that you intend to use are publicly available from the NCBI Sequence Read Archive, you can provide the identifiers of these libraries to \textbf{BRAKER3}. BRAKER3 will then automatically download the FASTQ files, align them to the genome with HiSat2, and convert the SAM file to the BAM format.

An example of an arbitrarily chosen Sequence Read Archive identifier is \texttt{ERR11837459} (paired-end Illumina RNA-Seq reads from the house spider).

Note that on some HPC environments, this may cause runtime limit issues if you provide a larger number of sequence identifiers. In such cases, it is recommended to download the data manually and provide the FASTQ (or aligned BAM) files to BRAKER3.

\subsubsubsection{FASTQ files \protect\includegraphics[width=0.8cm]{figs/braker3_optional.png}}

\textbf{BRAKER3} accepts as input a directory with FASTQ files of paired-end Illumina RNA-Seq data. The files must be named so that the first read of a pair has the suffix \texttt{\_1.fq} and the second read of a pair has the suffix \texttt{\_2.fq}. Both files in a pair must have the same prefix. BRAKER3 will then align the reads to the genome with HiSat2 and convert the SAM file to BAM format. Below is an example of two FASTQ formatted reads in two files:

\begin{exampleoutput}{reads\_1.fq}
@SEQ_ID_1/1
AGTCTGACGTGCTAGCTAGTCTGATCGATGCTAGCTAG
+
!''*((((***+))%%%++)(%%%%).1***-+*''))**55CCF>>>>>>CCCCCCC65
\end{exampleoutput}

\begin{exampleoutput}{reads\_2.fq}
@SEQ_ID_1/2
TGCATGCTAGCTAGCATCGATCAGACTAGCTAGCACGTCAGACT
+
!''*((((***+))%%%++)(%%%%).1***-+*''))**55CCF>>>>>>CCCCCCC65
\end{exampleoutput}

\subsubsection{Protein input for BRAKER2, BRAKER3, and Galba \protect\includegraphics[width=0.8cm]{figs/braker2.png}\protect\includegraphics[width=0.8cm]{figs/braker3.png}\protect\includegraphics[width=0.8cm]{figs/galba.png}}

\textbf{BRAKER3}, \textbf{BRAKER2}, and \textbf{Galba} accept protein input in FASTA format. However, the content requirements for the protein file differ between pipelines. Although BRAKER expects a large database of clade-specific proteins, Galba requires a small set of proteins from closely related species. For BRAKER, this set of proteins from closely related species can be concatenated into the larger database. Here is an example of protein FASTA format:

\begin{exampleoutput}{proteins.fa}
>103372_0:000000
VFSSVSCGRGNAGQTNYGMANSIMERICEKRAEEGLPGLAIQWGAVGDVGLVADMQEDNKELIIG
>103372_0:000001
SQQFLESIYHYNNCRSRKAYCGHRMLEQEMKVLTLPGRANQEKRNVIYEFMLEHLDMSAKSELVVKLTKYILRELCDENSIDVTTE
\end{exampleoutput}

Similarly to the genome file, the FASTA headers should be short and should not contain any special characters or white spaces. Sequences must be encoded in utf-8. The script \texttt{simplifyFastaHeaders.pl} can be used to generate a new protein file with simplified headers, along with a mapping table to cross-reference the original and updated headers.

\subsubsubsection{Large database of clade-specific proteins\protect\includegraphics[width=0.8cm]{figs/braker3.png}\protect\includegraphics[width=0.8cm]{figs/braker2.png}}

We recommend using OrthoDB partitions as input for \textbf{BRAKER3} and \textbf{BRAKER2}. OrthoDB is a comprehensive database of orthologous groups that covers a wide range of species. The database is partitioned into clade-specific databases, which can be downloaded from \url{https://bioinf.uni-greifswald.de/bioinf/partitioned_odb11/}. For example, to annotate the house spider or an insect, you can download the compressed file \texttt{Arthropoda.fa.gz}. The file can be decompressed with \texttt{gunzip} and used as input to BRAKER3 or BRAKER2.

\begin{bashinput}
gunzip Arthropoda.fa.gz
\end{bashinput}

This results in the extracted file \texttt{Arthropoda.fa} that can be used for BRAKER2 or BRAKER3.

\subsubsubsection{Small set of proteins from closely related species \protect\includegraphics[width=0.8cm]{figs/galba.png}\protect\includegraphics[width=0.8cm]{figs/braker3_optional.png}\protect\includegraphics[width=0.8cm]{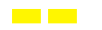}}

We recommend downloading annotated proteins for a hand full of closely related species as input for \textbf{Galba}. These proteins can also be combined with an OrthoDB partition for \textbf{BRAKER3} and \textbf{BRAKER2}. We use NCBI datasets (\url{https://www.ncbi.nlm.nih.gov/datasets/}) as an entry point to identify such protein sets. Enter the species name and click search. Click on the species and move to \texttt{Browse Taxonomy}. You can now see the number of available genomes displayed next to different taxonomic ranks. Pick the lowest rank that has >10 genomes available (you should aim for a minimum of 4 protein data sets). Click on the number of genomes. You will see a list of available genomes. Filter for genomes with an annotation, exclude atypical genomes. Select by clicking on the checkboxes of the genomes for which you want to download proteins files and click on \texttt{Download} $\rightarrow$ \texttt{Download Package}. Select \texttt{Protein (FASTA)}. The proteins will be downloaded into a zip archive. After extraction of the zip archive, you can concatenate the protein files into a single file with the \texttt{cat} command. We recommend that you simplify the FASTA headers with the \texttt{simplifyFastaHeaders.pl} script. In the following, we show an example of how to extract, concatenate, and simplify the headers of protein files of 13 related species for the house spider after download of the \texttt{ncbi\_dataset.zip} file:

\begin{bashinput}
unzip ncbi_dataset.zip
cat ncbi_dataset/data/*/protein.faa > proteins.fa
simplifyFastaHeaders.pl proteins.fa prot proteins_s.fa simplified.headers
\end{bashinput}

The file \texttt{proteins\_s.fa} can be used as input to Galba, or concatenated to an OrthoDB partition for BRAKER2 or BRAKER3. Here is an example command for such a concatenation process to prepare a protein file for BRAKER2 or BRAKER3:

\begin{bashinput}
cat proteins_s.fa Arthropoda.fa > all_proteins.fa
\end{bashinput}

The file \texttt{all\_proteins.fa} can be used as input for BRAKER2 or BRAKER3.

\subsubsection{Gene models in GTF format \protect\includegraphics[width=0.8cm]{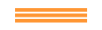}}\label{gtf}

TSEBRA, the Transcript Selector for BRAKER, is typically used to combine gene models from different sources. TSEBRA is compatible with AUGUSTUS-flavored GTF files produced by both BRAKER and Galba. The GTF format is a tab-separated file format that contains information about gene models. Below, you find an example of a GTF file with two alternative transcripts of one gene:

\begin{exampleoutput}{genes.gtf}
X1  AUGUSTUS      gene            107936  109553  .    +  .  g25
X1  AUGUSTUS      transcript      107936  109553  0.55 +  .  g25.t1
X1  AUGUSTUS      start_codon     107936  107938  .    +  0  transcript_id "g25.t1"; gene_id "g25";
X1  AUGUSTUS      CDS             107936  108168  1    +  0  transcript_id "g25.t1"; gene_id "g25";
X1  AUGUSTUS      exon            107936  108168  .    +  .  transcript_id "g25.t1"; gene_id "g25";
X1  AUGUSTUS      intron          108169  108601  1    +  .  transcript_id "g25.t1"; gene_id "g25";
X1  AUGUSTUS      CDS             108602  109005  0.93 +  1  transcript_id "g25.t1"; gene_id "g25";
X1  AUGUSTUS      exon            108602  109005  .    +  .  transcript_id "g25.t1"; gene_id "g25";
X1  AUGUSTUS      intron          109006  109476  0.55 +  .  transcript_id "g25.t1"; gene_id "g25";
X1  AUGUSTUS      CDS             109477  109553  0.55 +  2  transcript_id "g25.t1"; gene_id "g25";
X1  AUGUSTUS      exon            109477  109553  .    +  .  transcript_id "g25.t1"; gene_id "g25";
X1  AUGUSTUS      stop_codon      109551  109553  .    +  0  transcript_id "g25.t1"; gene_id "g25";
X1  GeneMark.hmm3 transcript      107936  109049  .    +  .  g25.t2
X1  GeneMark.hmm3 start_codon     107936  107938  .    +  0  transcript_id "g25.t2"; gene_id "g25";
X1  GeneMark.hmm3 CDS             107936  108168  .    +  0  transcript_id "g25.t2"; gene_id "g25";
X1  GeneMark.hmm3 exon            107936  108168  0    +  .  transcript_id "g25.t2"; gene_id "g25";
X1  GeneMark.hmm3 intron          108169  108601  .    +  1  transcript_id "g25.t2"; gene_id "g25";
X1  GeneMark.hmm3 CDS             108602  109049  .    +  1  transcript_id "g25.t2"; gene_id "g25";
X1  GeneMark.hmm3 exon            108602  109049  0    +  .  transcript_id "g25.t2"; gene_id "g25";
X1  GeneMark.hmm3 stop_codon      109047  109049  .    +  0  transcript_id "g25.t2"; gene_id "g25";
\end{exampleoutput}

The first column contains the name of the genomic sequence, the second column contains the source of the gene model, the third column contains the feature type, the fourth and fifth columns contain the start and end positions of the feature, the sixth column contains a score if available, the seventh column contains the strand, the eighth column contains the phase if applicable, and the ninth column contains attributes. The attributes are separated by a semicolon and contain key-value pairs (except for gene lines that contain only the gene identifier). Key-value pairs are separated by a space. The key-value pairs contain information about the gene model, in particular the gene id and the transcript id.

\subsection{Example files}

Both BRAKER and Galba provide small example files (from \textit{Arabidospis thaliana}) that can be used to test the pipelines. The example files are located in the \texttt{/opt/BRAKER/example} or \texttt{/opt/Galba/example} directories of the respective containers, as well as in the GitHub repositories. The example files do not allow testing of all pipeline modes. In addition, we provide full genome-scale example files for the species \textit{Drosophila melanogaster} at \url{https://bioinf.uni-greifswald.de/bioinf/downloads/braker/data}. 

\begin{table}[h!]
	\begin{adjustbox}{max width=\textwidth}
	\begin{tabular}{lllrlr}
	\hline
	Line & File category & Toy file & Size & \textit{Drosophila melanogaster} file & Size \\
	\hline
	{\includegraphics[width=0.8cm]{figs/braker1.png}\includegraphics[width=0.8cm]{figs/braker2.png}\includegraphics[width=0.8cm]{figs/braker3.png}\includegraphics[width=0.8cm]{figs/galba.png}} & Genome & \href{https://raw.githubusercontent.com/Gaius-Augustus/BRAKER/master/example/genome.fa}{genome.fa} & 933 KB & \href{https://bioinf.uni-greifswald.de/bioinf/braker/data/genome.fa.gz}{genome.fa.gz} & 134 MB\\
	{\includegraphics[width=0.8cm]{figs/braker1_optional.png}\includegraphics[width=0.8cm]{figs/braker3.png}} & Short read BAM & \href{http://bioinf.uni-greifswald.de/augustus/datasets/RNAseq.bam}{RNAseq.bam} & 164 MB & \href{https://bioinf.uni-greifswald.de/bioinf/braker/data/rnaseq.bam}{rnaseq.bam} & 1.9 GB\\
	{\includegraphics[width=0.8cm]{figs/braker3_external.png}} & Long read BAM & -- & -- & \href{http://bioinf.uni-greifswald.de/augustus/datasets/isoseq.bam}{isoseq.bam} & 49 MB\\
	{\includegraphics[width=0.8cm]{figs/braker1.png}} & Hints & \href{https://raw.githubusercontent.com/Gaius-Augustus/BRAKER/master/example/RNAseq.hints}{RNAseq.hints} & 51 KB & \href{http://bioinf.uni-greifswald.de/augustus/datasets/hints.gff.gz}{hints.gff.gz} & 4 MB\\
	{\includegraphics[width=0.8cm]{figs/braker3_optional.png}} & SRA IDs & -- & -- & SRR19416937 & -- \\
	{\includegraphics[width=0.8cm]{figs/braker3_optional.png}} & FASTQ & -- & -- & \href{https://bioinf.uni-greifswald.de/bioinf/braker/data/file1_1.fastq.gz}{file1\_1.fastq.gz} & 9.2 GB\\
	 &       &-- & --&\href{https://bioinf.uni-greifswald.de/bioinf/braker/data/file1_2.fastq.gz}{file1\_2.fastq.gz} & 9.2 GB\\
	{\includegraphics[width=0.8cm]{figs/braker2.png}\includegraphics[width=0.8cm]{figs/braker3.png}} & Large protein database & \href{https://raw.githubusercontent.com/Gaius-Augustus/BRAKER/master/example/proteins.fa}{proteins.fa} & 3 MB & \href{https://bioinf.uni-greifswald.de/bioinf/partitioned_odb11/Arthropoda.fa.gz}{Arthropoda.fa.gz} & 2.2 GB\\
	{\includegraphics[width=0.8cm]{figs/galba.png}\includegraphics[width=0.8cm]{figs/braker2_optional.png}\includegraphics[width=0.8cm]{figs/braker3_optional.png}} & Proteins of close relatives & \href{https://raw.githubusercontent.com/Gaius-Augustus/GALBA/main/example/proteins.fa}{proteins.fa} & 116 KB &\href{https://bioinf.uni-greifswald.de/bioinf/braker/data/file1_1.fastq.gz}{proteins.fa.gz}  & 85 MB\\
	\hline
	\end{tabular}
\end{adjustbox}
\caption{Example files for testing BRAKER and Galba. For the fly data set, the file sizes of the extracted files are shown; i.e.,~after downloading, you have to unpack the files ending in \texttt{*.gz} with the \texttt{gunzip} tool. Afterward, you will see the final file sizes. The preparation of the fruit fly genome is described in \cite{bruuna2021braker2}. The short read RNA-Seq data sets for the fly were generated from library SRR19416937. The Iso-Seq file corresponds to data set 6 from \cite{krivzanovic2018evaluation}. Note that none of the provided data sets were designed to measure accuracy. For measuring accuracy, one would exclude the target species itself from the protein data set and choose more transcriptome data (for both short and long reads).\label{test_data}}
\end{table}
Full example command lines (and runtimes) with both the toy data as well as the fly data, are provided in the Appendix. In the main book chapter, we use placeholders to describe file names.

\section{Gene prediction, step by step} \label{methods}

\begin{figure}[h!]
	\begin{center}
		\includegraphics[width=0.5\linewidth]{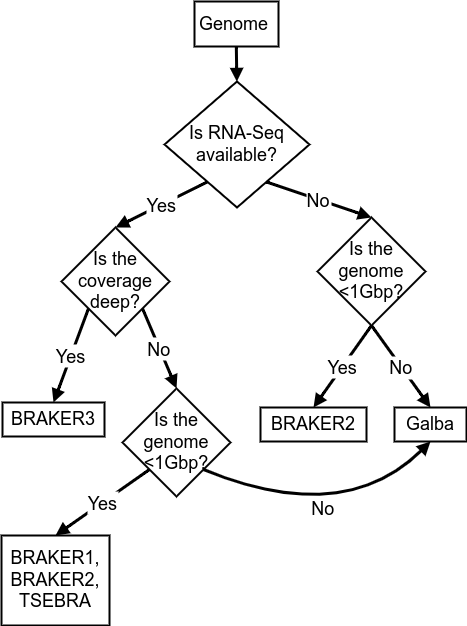}
		\caption{Decision scheme  for picking a suitable pipeline out of BRAKER3, BRAKER2, BRAKER1 (in combination with BRAKER2 and TSEBRA), and Galba. \label{decision}}
	\end{center}
\end{figure}

Running BRAKER and Galba should be intuitive, since they are designed as one-stop solutions for gene prediction. However, both pipelines are complex and offer many options. We will guide you through the most common use cases. Figure \ref{decision} provides a recommendation on which pipeline or pipeline combination to choose under what circumstances. In this book chapter, we begin describing the use case of BRAKER3, which is the pipeline that will in most cases achieve the best accuracy. We then move to BRAKER2 and Galba, which can be applied in the absence of transcriptome data. We also demonstrate how to run BRAKER1 with only transcriptome data, with the goal of combining such a gene set with a BRAKER2 gene set using TSEBRA, and we show how to run BRAKER3 with PacBio HiFi Iso-Seq transcriptome data. However, before we get started, it is necessary to understand how to make directories and files visible to the container.

\subsection{Making directories visible to the container}

The most common use case will involve Singularity on a Unix system since this does not require root permissions. To mount any directory to your Singularity container, 
use the option \texttt{-B} followed by the path to the directory, followed by a colon (\texttt{:}), followed by the path to the directory that the container shall use. The following example shows how to mount the current working directory to the BRAKER3 container Singularity file:

\begin{bashinput}
singularity exec -B ${PWD}:${PWD} braker3.sif /bin/bash
\end{bashinput}

This makes the current working directory where you are located visible to the container at the same location. You can now execute the BRAKER3 commands from within the container and see this directory and its contents. The user home directory should be mounted by default. You can exit the Bash that is invoked in the above example with \texttt{exit}.

If you prefer to use Docker, you can mount directories as follows:

\begin{bashinput}
sudo docker run -v ${PWD}:${PWD} -it teambraker/braker3:latest /bin/bash
\end{bashinput}

It is important to mount all directories that contain input files and that you want to write output to. You can easily mount more than one directory. If you do not mount a directory, the container will not be able to see it. In the following examples, we always assume that all our files are located in the current working directory and that we want to write to that same directory. We also assume that you prefer to use Singularity. If you prefer to use Docker, you can easily adapt the commands.

Note that in Singularity, the user's home directory is mounted by default, whereas in Docker it is not. To run BRAKER and Galba with Singularity, this is important because when AUGUSTUS trains parameters for a novel species, these parameter files need to be stored in a location called \texttt{\$AUGUSTUS\_CONFIG\_PATH}. The Singularity image is static and thus none of the pipelines will be able to write these parameter files into the Singularity image. Therefore, both Galba and BRAKER copy the entire contents of \texttt{\$AUGUSTUS\_CONFIG\_PATH} to the user's home directory at \texttt{$\sim$/.augustus}. If you run BRAKER or Galba from Docker, this is not a problem, since Docker can write into the container.
\subsection{Running BRAKER3}

BRAKER3 is the most accurate gene prediction pipeline described in this chapter. In the following, we describe in detail how to run BRAKER3 with a genome file, a protein database file, and SRA IDs as transcriptome input. Subsequently, you will find shorter sections that describe alternative ways to provide transcriptome data to BRAKER3.

\subsubsection{BRAKER3 with a protein database and SRA IDs \protect\includegraphics[width=0.8cm]{figs/braker3_optional.png} \protect\includegraphics[width=0.8cm]{figs/braker3.png}}

A standard run of BRAKER3 requires a genome file in FASTA format, a protein file in FASTA format, and some kind of (short read) transcriptome input. The following example shows how to run BRAKER3 with the example genome file \texttt{genome.fa}, the example protein file \texttt{proteins.fa}, and the Sequence Read Archive identifiers of two RNA-Seq libraries:

\begin{bashinput}
T=72 # specify number of threads
singularity exec -B ${PWD}:${PWD} braker3.sif braker.pl --species=yourSpecies --genome=genome.fa \
  --prot_seq=proteins.fa --rnaseq_sets_ids=SRR12345678,SRR12345679 --workingdir=braker3 \
  --threads=${T} --busco_lineage=arthropoda_odb10
\end{bashinput}

Understanding this command:

\begin{itemize}
\item \texttt{singularity exec -B \${PWD}:\${PWD} braker3.sif} is the command to execute the BRAKER3 container. The option \texttt{-B \${PWD}:\${PWD}} mounts the current working directory on the container. This is important because the container will not be able to see any files that are not mounted on it.
\item \texttt{braker.pl} is the BRAKER3 command-line tool that we call. 
\item \texttt{-{}-species=yourSpecies} is an optional argument used to specify the species name that will be used to store the AUGUSTUS parameters in \texttt{\${AUGUSTUS\_CONFIG\_PATH}}. If you do not provide a species name, BRAKER3 will generate a nondescriptive species name. Since running AUGUSTUS training is computationally expensive, we recommend using a species name that is descriptive for your species. The species name should not contain special characters or white spaces. The species name must be unique and it cannot already exist in the \${AUGUSTUS\_CONFIG\_PATH} directory. If you run BRAKER (or Galba) with the same species name a second times, the pipeline will die.
\item \texttt{-{}-genome=genome.fa} is the argument to specify the genome file in FASTA format. This file should be soft-masked for repeats.
\item \texttt{-{}-prot\_seq=proteins.fa} is the argument for specifying the protein file in FASTA format. This file should contain a large database of clade-specific proteins. Commonly used OrthoDB partitions are available at \url{https://bioinf.uni-greifswald.de/bioinf/partitioned_odb11/}.
\item \texttt{-{}-rnaseq\_sets\_ids=SRR12345678,SRR12345679} is the argument for specifying the sequence read archive identifiers of the RNA-Seq libraries. You can provide as many identifiers as you want. If you provide more than one identifier, you have to separate them by a comma. If you provide a large number of identifiers, you may run into run-time limit issues in some HPC environments. There are also other options to provide RNA-Seq data to BRAKER3 (see the next sections).
\item \texttt{-{}-workingdir=braker3} is an optional argument to specify the output directory. If you do not provide a working directory, BRAKER3 will create a directory called \texttt{braker} in the current working directory.
\item \texttt{-{}-threads=\${T}} is an optional argument to specify the number of threads. If you do not provide a number of threads, BRAKER3 will use 1 thread. We recommend using a minimum of 8 threads. Using a massive number of threads may cause problems because BRAKER3 uses data parallelization in many steps. Massive data splitting leads to input-output problems and may slow down the pipeline. We recommend using a maximum of 72 threads.
\item \texttt{-{}-busco\_lineage=arthropoda\_odb10} is an optional but recommended argument. It is used to specify the BUSCO lineage for a compleasm assessment of both genome and predicted protein sets. The lineage \texttt{arthropoda\_odb10} is a good choice for arthropods.
\end{itemize}

If you need the output in GFF3 format, you can add the option \texttt{-{}-gff3} that is available in all variants of BRAKER and also in Galba. 

Option \texttt{-{}-AUGUSTUS\_CONFIG\_PATH=/some/directory} allows you to specify a different local directory that contains AUGUSTUS parameters and is writable. This is useful if you run BRAKER on a shared file system and you want to avoid BRAKER writing to the home directory. Note that you have to mount the location of that directory on the container with the \texttt{-B} option.

It is possible to skip a runtime-consuming step of training AUGUSTUS with the argument \texttt{-{}-skipOptimize}. Optimizing AUGUSTUS parameters improves accuracy by a few percent points but may run for many hours. If you are in a hurry, and if accuracy is less important, you can skip this step.

Invoked like this, BRAKER3 will perform the following steps:
\begin{enumerate}
\item Download the RNA-Seq data from the Sequence Read Archive.
\item Align the RNA-Seq data to the genome with HiSat2.
\item Convert the SAM file to BAM format.
\item Run self-training GeneMark-ETP to predict genes from the RNA-Seq and protein data. Internally, GeneMark-ETP performs a StringTie assembly, calls genes in that assembly with GeneMarkS-T, de-noise these predictions with filtering based on DIAMOND hits, and invokes ProtHint.
\item Train AUGUSTUS with selected GeneMark-ETP gene predictions.
\item Run compleasm to generate evidence of BUSCOs for AUGUSTUS.
\item Run AUGUSTUS with the trained parameters and RNA-Seq, protein data, and BUSCO evidence.
\item Run TSEBRA to combine the AUGUSTUS gene predictions with the GeneMark-ETP gene predictions.
\item Run compleasm to assess the completeness of the TSEBRA gene set, the GeneMark-ETP gene set, and the AUGUSTUS gene set. Depending on the results of this assessment, it may rerun TSEBRA with different parameters to produce an improved gene set.
\end{enumerate}

\begin{figure}[h!]
	\begin{center}
		\includegraphics[width=\linewidth]{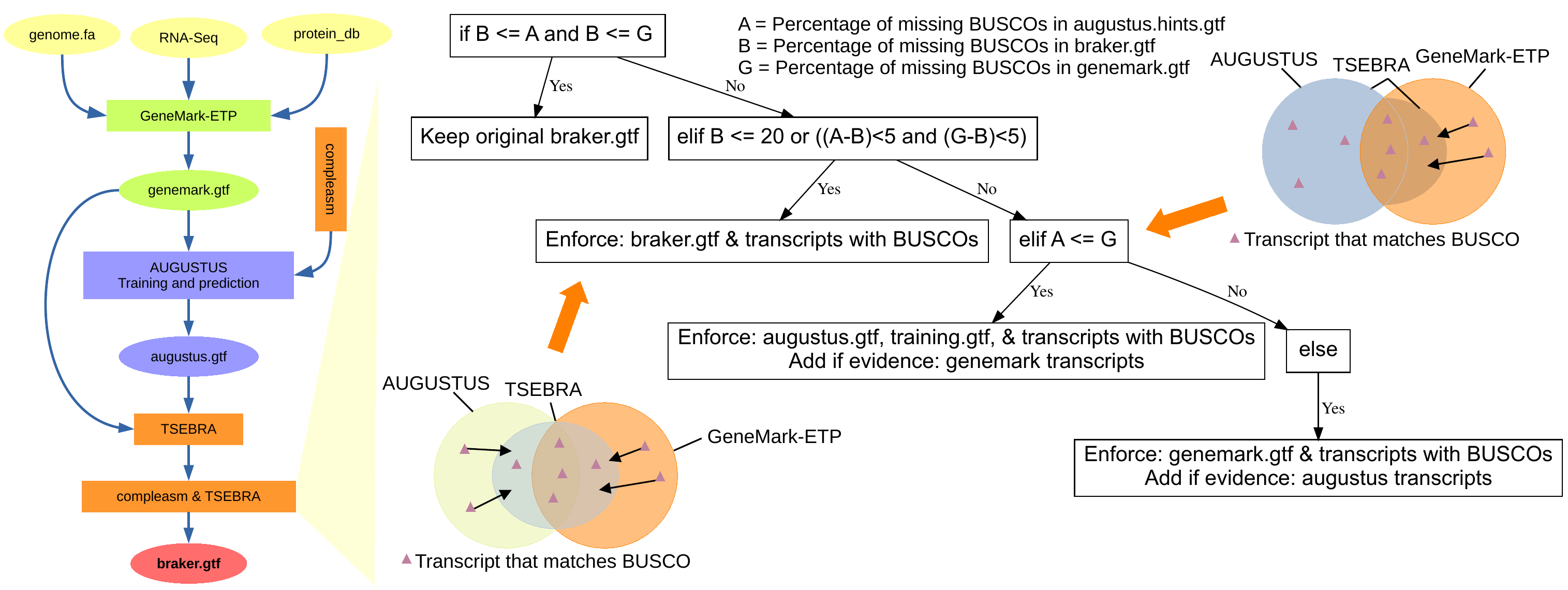}
		\caption{If invoked with the option \texttt{-{}-busco\_lineage}, BRAKER runs compleasm \cite{huang2023compleasm}  on genome level to generate strong evidence for clade-specific marker genes that will be used by AUGUSTUS during gene prediction. 
		Furthermore, compleasm is used to assess the amount of missing BUSCOs in the initial TSEBRA gene set, in the GeneMark gene set, and in the AUGUSTUS gene set. Depending on the situation, a decision scheme with manually adapted thresholds is applied: if the initial TSEBRA output has the fewest missing BUSCOs, nothing happens. If one of the AUGUSTUS or GeneMark gene sets has the fewest missing BUSCOs, the following happens: If the original set of TSEBRA genes lacks fewer than 20\% BUSCOs, or if the difference between AUGUSTUS and TSEBRA, as well as between GeneMark and TSEBRA, is smaller than 5 percent points, then the initial TSEBRA gene set is kept. Furthermore, in this case, all GeneMark and AUGUSTUS transcripts that were not initially included in the TSEBRA gene set are added if they are similar to BUSCOs. If none of the above conditions are satisfied, it is determined whether AUGUSTUS or GeneMark has the lowest number of missing BUSCOs, and that gene set is kept. BUSCO-containing transcripts from the respective other gene set are added to it. \label{compleasm}}
	\end{center}
\end{figure}

\begin{figure}[h!]
	\begin{center}
		\includegraphics[width=\linewidth]{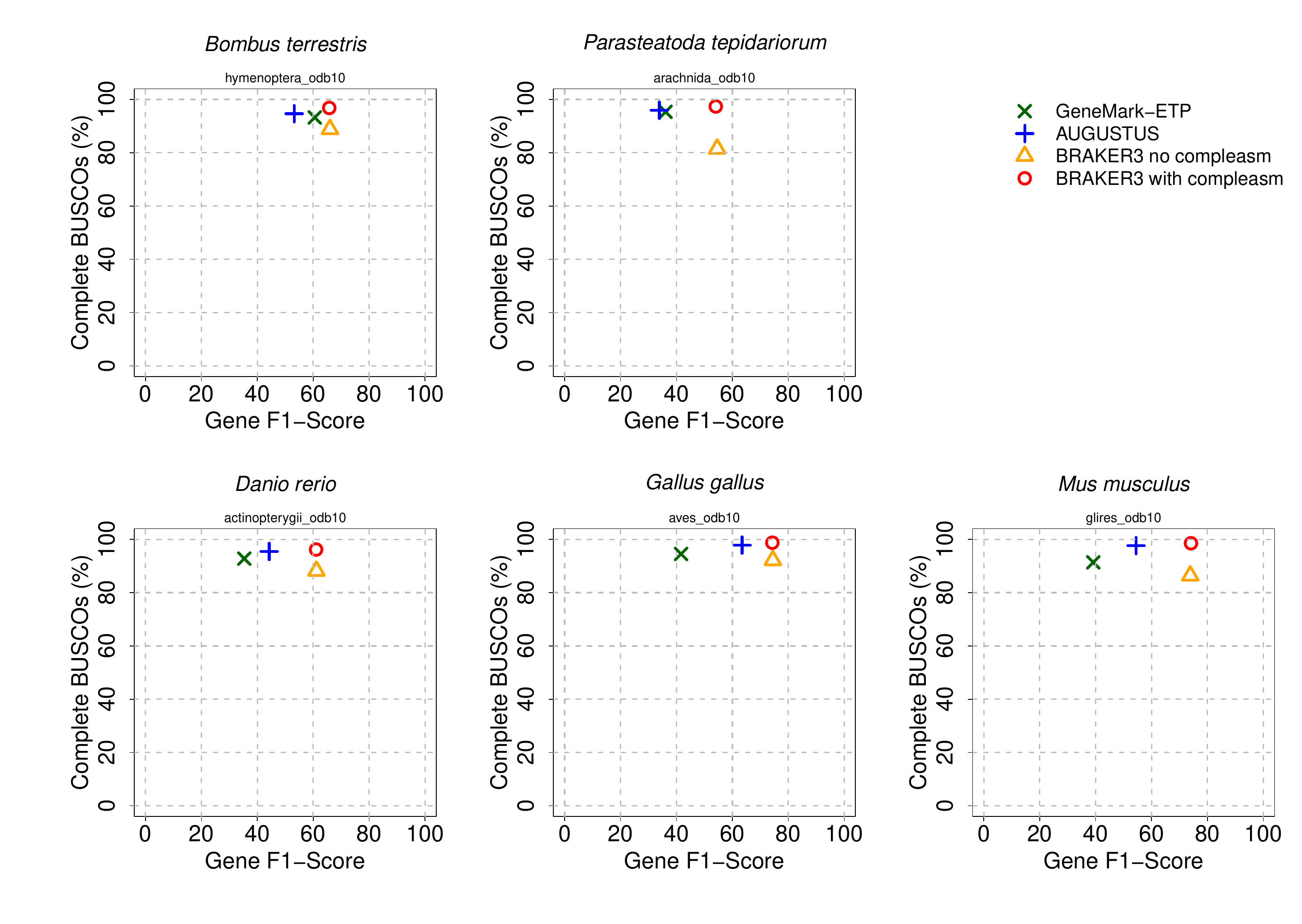}
		\caption{When the option \texttt{-{}-busco\_lineage} is used in BRAKER3, accuracy remains constant but BUSCO-completeness increases. The data used for this assessment is identical to the test data described in \cite{gabriel2023braker3}. The Gene F1-Score is here the harmonic mean of sensitivity ($\frac{TP}{TP+FN}$) and precision ($\frac{TP}{TP+FP}$), BUSCO completeness for the plot was determined by BUSCO \cite{BUSCO15} in protein mode.\label{compleasm-acc}}
	\end{center}
\end{figure}

The compleasm extension is a recent development that was added after the submission of the BRAKER3 manuscript \cite{gabriel2023braker3}. Compleasm was custom-extended by Neng Huang to assess the completeness of protein sets in addition to the completeness of genomes. We illustrate the decision scheme used in BRAKER in Figure \ref{compleasm}. BRAKER was extended in this way to ensure that the final gene sets have a large number of BUSCOs. Parameters were manually set to enhance accuracy in well-annotated genomes. This improvement is documented in Figure \ref{compleasm-acc}, which shows that for selected species with previously suboptimal BUSCO scores in the original TSEBRA gene set, invoking the \texttt{-{}-busco\_lineage} option leads to enhanced BUSCO completeness while maintaining constant accuracy. Note that the option works similarly for BRAKER1 and BRAKER2.

Command lines with the fly data set are provided in Appendix \ref{braker3sra}.

\subsubsection{BRAKER3 with a protein database and FASTQ files \protect\includegraphics[width=0.8cm]{figs/braker3_optional.png} \protect\includegraphics[width=0.8cm]{figs/braker3.png}}

If you do not have access to Sequence Read Archive identifiers, you can provide a directory with FASTQ files to BRAKER3. The following example shows how to run BRAKER3 with the example genome file \texttt{genome.fa}, the example protein file \texttt{proteins.fa}, and a directory with FASTQ files:

\begin{bashinput}
T=72 # specify number of threads
singularity exec -B ${PWD}:${PWD} braker3.sif braker.pl --species=yourSpecies --genome=genome.fa \
  --prot_seq=proteins.fa --rnaseq_sets_dir=/path/to/fastq/files/ --rnaseq_sets_ids=file1,file2 \ 
  --workingdir=braker3 --threads=${T} --busco_lineage=arthropoda_odb10 
\end{bashinput}

The option \texttt{-{}-rnaseq\_sets\_dir} is used to specify the directory with the FASTQ files. 

The option \texttt{-{}-rnaseq\_sets\_ids} is used to specify the name stems of the FASTQ files. Here, we assume that the files 
\begin{itemize}
\item \texttt{/path/to/fastq/files/file1\_1.fastq}, 
\item \texttt{/path/to/fastq/files/file1\_2.fastq},
\item \texttt{/path/to/fastq/files/file2\_1.fastq}, 
\item and \texttt{/path/to/fastq/files/file2\_2.fastq} 
\end{itemize}
are available and contain short paired end transcriptome reads in FASTQ format. The name stems are \texttt{file1} and \texttt{file2}. The name stems are used to identify the paired-end FASTQ files, they must be identical for both files in a pair and must not contain any special characters or white spaces. They are provided as values for the option \texttt{-{}-rnaseq\_sets\_ids}. You can provide an unlimited number of FASTQ files; multiple file names need to be separated with a comma. Providing a large number of files may lead to runtime limit issues in some HPC environments. There are also other options to provide RNA-Seq data to BRAKER3 (see the next sections).

BRAKER3 will execute the same steps as in the previous example. The only difference is that it will not download the RNA-Seq data from the Sequence Read Archive. Instead, it will use the FASTQ files that you provide.

Command lines with the fly data set are provided in Appendix \ref{braker3fastq}.

\subsubsection{BRAKER3 with a protein database and a BAM file \protect\includegraphics[width=0.8cm]{figs/braker3.png}}

Alternatively to SRA IDs or FASTQ files, you can provide one or several externally prepared BAM files with spliced alignments to BRAKER3. We recommend merging and sorting BAM files before evoking BRAKER since this is a runtime-intensive step that sometimes fails due to input output problems. The following example shows how to run BRAKER3 with the example genome file \texttt{genome.fa}, the example protein file \texttt{proteins.fa}, and a BAM file:

\begin{bashinput}
T=72 # specify number of threads
singularity exec -B ${PWD}:${PWD} braker3.sif braker.pl --species=yourSpecies --genome=genome.fa \
  --prot_seq=proteins.fa --bam=rnaseq.bam --workingdir=braker3 --threads=${T} \ 
  --busco_lineage=arthropoda_odb10
\end{bashinput}

The option \texttt{-{}-bam} is used to specify the BAM file. Here, we assume that the file \texttt{rnaseq.bam} is available and contains spliced alignments of short paired end transcriptome reads to the genome. It is possible to provide several comma-separated BAM files. Note that BRAKER3 is developed with HiSat2 BAM files. Not all BAM files of all possible spliced aligners are compatible with BRAKER.

BRAKER3 will execute the same steps as in the previous examples. The only difference is that it will use the BAM file that you provide.

It is also possible to provide BAM files that were previously generated for each RNA-Seq library in the folder provided at \texttt{-{}-rnaseq\_sets\_dir}. Example:

\begin{bashinput}
T=72 # specify number of threads
singularity exec -B ${PWD}:${PWD} braker3.sif braker.pl --species=yourSpecies --genome=genome.fa \
  --prot_seq=proteins.fa --rnaseq_sets_dir=/path/to/bam/files/ --rnaseq_sets_ids=test1,test2 \ 
  --workingdir=braker3 --threads=${T} --busco_lineage=arthropoda_odb10
\end{bashinput}

BRAKER will automatically detect that these ID(s) are BAM and not FASTQ file(s) if the BAM files exist, in this example \texttt{/path/to/bam/files/test1.bam} and \texttt{/path/to/bam/files/test2.bam}.

Command lines with the toy data set are provided in Appendix \ref{braker3bam}, command lines with the fly data sets are provided in Appendix \ref{braker3bamfly}.

\subsubsection{BRAKER3 with a BAM file from Iso-Seq data \protect\includegraphics[width=0.8cm]{figs/braker3_external.png}\protect\includegraphics[width=0.8cm]{figs/braker3.png}}

If you have Iso-Seq data, you can provide a BAM file with spliced alignments to BRAKER3. Note that you have to use a container with a different tag for this (see Section \ref{isoseq}). The following example shows how to run BRAKER3 with the example genome file \texttt{genome.fa}, the example protein file \texttt{proteins.fa}, and a BAM file:

\begin{bashinput}
T=72 # specify number of threads
singularity exec -B ${PWD}:${PWD} braker3_isoseq.sif braker.pl --species=yourSpecies \
  --genome=genome.fa --prot_seq=proteins.fa --bam=isoseq.bam --workingdir=braker3 \
  --threads=${T} --busco_lineage=arthropoda_odb10
\end{bashinput}

Do not mix short- and long-read transcriptome data with the current versions of BRAKER3 and GeneMark-ETP. If you have both types of data, we recommend running BRAKER3 with the short read data first and then to run BRAKER3 with the long read data. You can then use TSEBRA to combine the gene sets.

\begin{figure}[h!]
	\begin{center}
		\includegraphics[width=\linewidth]{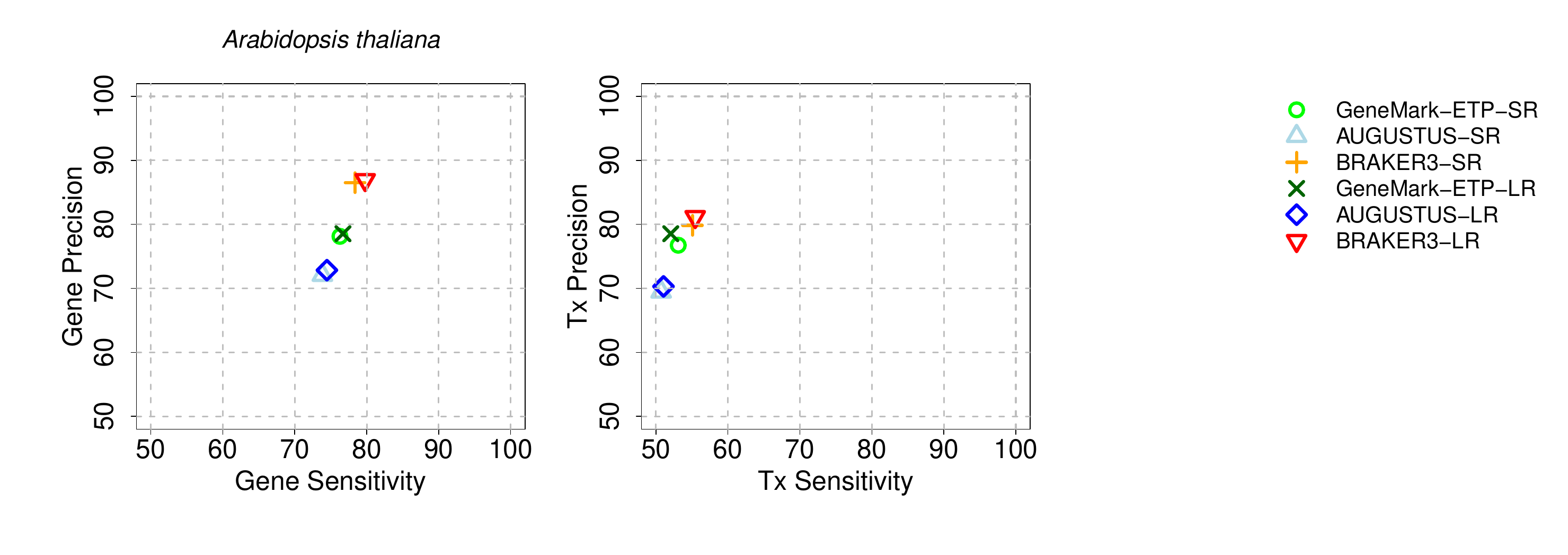}
		\caption{Accuracy of BRAKER3 and its underlying gene finders with short read RNA-Seq data (-SR) and with long read Iso-Seq (-LR) from \cite{zhang2022high}. (No \texttt{-{}-busco\_lineage} was provided.) The short read RNA-Seq, genomic, protein database, and reference annotation data used for this assessment is identical to the test data described in \cite{gabriel2023braker3}. Sensitivity is here defined as $\frac{TP}{TP+FN}$ and precision as $\frac{TP}{TP+FP}$. Accuracy is reported on gene and transcript (Tx) level.\label{isoseq-acc}}
	\end{center}
\end{figure}

BRAKER3 has not been extensively tested with Iso-Seq input because it has proven difficult to obtain a sufficient amount of high quality reads for reference organisms that allow accuracy benchmarking. However, we provide exemplary accuracy results for the genome of \textit{Arabidopsis thaliana} for which Zhang \textit{et al.} (2022) provide 11 GB of Iso-Seq reads (7,604,981 reads, 96\% alignment rate to the genome when mapped with minimap2 as described above) \cite{zhang2022high}. When comparing to accuracy results obtained with short read RNA-Seq data, we find that BRAKER3 with Iso-Seq data is slightly more accurate on gene and transcript level (Figure \ref{isoseq-acc}). Please note that such results are only achievable if you provide a large amount of high quality Iso-Seq data to BRAKER3. (Regrettably, we were unable to find a comparable data set for an insect reference genome at the time of writing this chapter.)

Command lines with the fly data set are provided in Appendix \ref{braker3isoseqfly}.

\subsection{Running BRAKER2 with a protein database \protect\includegraphics[width=0.8cm]{figs/braker2.png}}

BRAKER2 is a pipeline that can be used in the absence of transcriptome data, operating on similar principles as BRAKER3. Be aware that BRAKER2 and its underlying gene prediction tool GeneMark-EP require a certain degree of sequence redundancy in the protein database. On average, you should have a minimum of four homologous proteins per gene with evidence. OrthoDB partitions satisfy this criterion and may often be sufficient. For clades with few representatives in OrthoDB, you may want to concatenate additional proteomes of closely related species to an OrthoDB partition. GeneMark-EP, a critical component of BRAKER2, does not perform well in very large and repeat-rich genomes. In such cases, we recommend using Galba instead.

The following example shows how to run BRAKER2 with the example genome file \texttt{genome.fa} of a small or medium sized genome (<1Gbp) and the example protein database file \texttt{proteins.fa}:

\begin{bashinput}
T=72 # specify number of threads
singularity exec -B ${PWD}:${PWD} braker3.sif braker.pl --species=yourSpecies --genome=genome.fa \
  --prot_seq=proteins.fa --workingdir=braker2 --threads=${T} --busco_lineage=arthropoda_odb10
\end{bashinput}

BRAKER2 will execute the following steps:
\begin{enumerate}
\item Run self-training GeneMark-EP (including the ProtHint pipeline for generating protein evidence) to predict genes, partially with support of protein evidence.
\item Train AUGUSTUS with selected GeneMark-EP gene predictions.
\item Run compleasm to generate evidence of BUSCOs for AUGUSTUS.
\item Run AUGUSTUS with the trained parameters and the protein data.
\item Feed the AUGUSTUS gene predictions back to ProtHint to generate more evidence.
\item Run AUGUSTUS with the trained parameters and the protein data and the additional evidence from ProtHint.
\item Run TSEBRA to combine the set of AUGUSTUS genes with evidence-supported GeneMark genes.
\item Run compleasm to assess the completeness of the TSEBRA gene set, the GeneMark-EP gene set, and the AUGUSTUS gene set. Depending on the results of this assessment, it may re-run TSEBRA with different parameters to produce an improved gene set.
\end{enumerate}

Command lines with the toy data set are provided in Appendix \ref{braker2toy}, command lines with the fly data set are provided in Appendix \ref{braker2fly}.

\subsection{Running Galba with proteins of closely related species \protect\includegraphics[width=0.8cm]{figs/galba.png}}

Galba is another pipeline that can be used in the absence of transcriptome data. Galba is based on principles similar to BRAKER2, but does not rely on GeneMark-EP. Instead, it employs the miniprot direct protein-to-genome spliced alignment tool to generate training genes for AUGUSTUS. It is important to note that Galba also benefits from a certain degree of sequence redundancy in the protein database; however, this redundancy is not strictly required. Galba can be successfully completed with a reference proteome of a single related species. In contrast to BRAKER2, providing a huge database of proteins, such as an OrthoDB partition, is not advisable. Although miniprot is fast, it is much slower than ProtHint with a large input. Furthermore, providing an OrthoDB partition alone may cause Galba to remove too many predictions in the final filtering step. The following example shows how to run Galba with the example genome file \texttt{genome.fa} and the example protein file \texttt{proteins.fa}:

\begin{bashinput}
T=72 # specify number of threads
singularity exec -B ${PWD}:${PWD} galba.sif galba.pl --species=yourSpecies --genome=genome.fa \
  --prot_seq=proteins.fa --workingdir=galba --threads=${T}
\end{bashinput}

Galba will execute the following steps:
\begin{enumerate}
\item Run miniprot to align protein sequences of a few closely related species to the genome.
\item Run miniprothint to filter alignments for quality and identify training genes that are then used to train AUGUSTUS.
\item Run AUGUSTUS with the filtered protein evidence.
\item Run the initial predictions through a quality filter and use them for a second training iteration.
\item Run AUGUSTUS with the second trained parameters and the filtered protein evidence.
\item Filter the predictions by searching for DIAMOND hits in the input protein data set to decrease false positive predictions.
\end{enumerate}

The command line options of BRAKER and GALBA are highly similar since GALBA was developed from the BRAKER pipeline code base.

Command lines with the toy data set are provided in Appendix \ref{galbatoy}, command lines with the fly data set are provided in Appendix \ref{galbafly}.

\subsection{Running BRAKER1}

BRAKER1 was historically the first version of the BRAKER pipeline. It does not use protein data, but relies on information of spliced alignments from transcriptomics only. Thus, BRAKER1 exploits transcriptome information to a lesser extent than BRAKER3. Similarly to BRAKER2, BRAKER1 does not achieve high accuracy in large genomes (>1Gbp) and repeat-rich genomes. In most cases, we recommend using BRAKER3 because a protein database will always be available when transcriptome data is also available. However, BRAKER3 needs sufficient transcriptome sequencing depth to perform StringTie2 assembly and call genes with GeneMarkS-T. In some cases, the depth of transcriptome sequencing may be insufficient for BRAKER3. In such cases, BRAKER1 may still be a good choice, in particular if these obtained predictions are later combined with BRAKER2 predictions using TSEBRA. In the following, we first show how to run BRAKER1 with an externally prepared BAM file containing short read transcriptome-spliced alignments. We then demonstrate how to run BRAKER1 with an externally prepared hints file.

Note that BRAKER1 relies on coverage information for splice sites. Therefore, it is not suitable to use the splice junctions obtained from mapping a single-species assembled transcriptome against the genome. Instead, it is important to use splice junctions obtained from mapping short read transcriptome data against the genome.

\subsubsection{BRAKER1 with a BAM file \protect\includegraphics[width=0.8cm]{figs/braker1_optional.png} \protect\includegraphics[width=0.8cm]{figs/braker1.png}}

The following example shows how to run BRAKER1 with the example genome file \texttt{genome.fa} and the example RNA-Seq BAM file:

\begin{bashinput}
T=72 # specify number of threads
singularity exec -B ${PWD}:${PWD} braker3.sif braker.pl --species=yourSpecies --genome=genome.fa \
  --bam=rnaseq.bam --workingdir=braker1 --threads=${T} --busco_lineage=arthropoda_odb10
\end{bashinput}

Similarly to BRAKER3, here it is also possible to provide more than one BAM file in a comma-separated fashion. BRAKER1 will execute the following steps:

\begin{enumerate}
\item Run the self-training GeneMark-ET to predict genes from the RNA-Seq data.
\item Train AUGUSTUS with selected GeneMark-ET gene predictions.
\item Run AUGUSTUS with the trained parameters and the RNA-Seq data.
\item Run TSEBRA to combine the set of AUGUSTUS genes with evidence-supported GeneMark genes.
\item Run compleasm to assess the completeness of the TSEBRA gene set, the GeneMark-ET gene set, and the AUGUSTUS gene set. Depending on the results of this assessment, it may re-run TSEBRA with different parameters to produce an improved gene set.
\end{enumerate}

Command lines with the toy data set are provided in Appendix \ref{braker1toy}, command lines with the fly data set are provided in Appendix \ref{braker1fly}.

\subsubsection{BRAKER1 with a hints file \protect\includegraphics[width=0.8cm]{figs/braker1.png}}

If you have a hints file, you can provide it to BRAKER1. The hints file should be the AUGUSTUS-flavored GFF hints format. The following example shows how to run BRAKER1 with the example genome file \texttt{genome.fa} and the example hints file \texttt{hints.gff}:

\begin{bashinput}
T=72 # specify number of threads
singularity exec -B ${PWD}:${PWD} braker3.sif braker.pl --species=yourSpecies --genome=genome.fa \
  --hints=hints.gff --workingdir=braker1 --threads=${T} --busco_lineage=arthropoda_odb10
\end{bashinput}

It is also possible to provide more than one hints file in a comma separated fashion.

Command lines with the toy data set are provided in Appendix \ref{braker1toyhints}, command lines with the fly data set are provided in Appendix \ref{braker1hintsfly}.

\subsection{Output of BRAKER}

All BRAKER pipelines write the following output files to the working directory:

\begin{itemize}
	\item \texttt{braker.log}
	\item \texttt{braker.gtf}
	\item \texttt{braker.codingseq}
	\item \texttt{braker.aa}
	\item \texttt{hintsfile.gff}
\end{itemize}

The \texttt{braker.log} file contains the log of the BRAKER run. It may contain possible error messages if a BRAKER run terminates prematurely. The \texttt{braker.gtf} file contains the gene predictions in GTF format (see section \ref{gtf} on page \pageref{gtf}). Note that the final gene models may originate from AUGUSTUS and GeneMark. The \texttt{braker.codingseq} file contains the coding sequences of the gene predictions in FASTA format. The \texttt{braker.aa} file contains the amino acid sequences of the gene predictions in FASTA format. The \texttt{hintsfile.gff} file contains the hints that were used for gene prediction in GFF format. The hints may be useful for running TSEBRA post hoc.

\subsection{Output of Galba}

Galba writes the following output files to the working directory:

\begin{itemize}
	\item \texttt{GALBA.log}
	\item \texttt{galba.gtf}
	\item \texttt{galba.codingseq}
	\item \texttt{galba.aa}
	\item \texttt{hintsfile.gff}
\end{itemize}

The \texttt{GALBA.log} file contains the GALBA run log. It may contain possible error messages if a GALBA run terminated prematurely. The \texttt{galba.gtf} file contains the gene predictions in GTF format (see Section \ref{gtf} on page \pageref{gtf}). The \texttt{galba.codingseq} file contains the coding sequences of the gene predictions in FASTA format. The \texttt{galba.aa} file contains the amino acid sequences of the gene predictions in FASTA format. The \texttt{hintsfile.gff} file contains the hints that were used for gene prediction in GFF format. The hints may be useful for running TSEBRA.

\subsection{Running TSEBRA \protect\includegraphics[width=0.8cm]{figs/tsebra.png}}

BRAKER already internally executes TSEBRA to combine the underlying GeneMark and AUGUSTUS gene sets. However, sometimes you may want to combine the BRAKER (or Galba) gene sets with each other. The most common scenario will be to combine the results of a BRAKER1 run with the results of a BRAKER2 run in case the RNA-Seq coverage was not sufficient to successfully run BRAKER3. 

By default, TSEBRA will first combine transcripts from two or more files into a nonredundant set, and subsequently use the evidence from the hints file to decide which transcripts to keep and which transcripts to drop. TSEBRA has two separate strategies for handling single-exon genes. In large genomes (>1Gbp), only single-exon genes supported by hints will be kept. In small and medium-sized genomes, all single-exon genes will be kept.

We see two typical use-case scenarios for running TSEBRA: (1) combining gene sets and deciding on keeping transcripts using the evidence, and (2) combining gene sets while one gene set is enforced. The following example shows how to run TSEBRA with the example gene set \texttt{BRAKER1/braker1.gtf} and the example gene set \texttt{BRAKER2/braker2.gtf} for an evidence-based transcript selection:

\begin{bashinput}
singularity exec -B ${PWD}:${PWD} braker3.sif tsebra.py \
  -g BRAKER1/braker1.gtf,BRAKER2/braker2.gtf \
  -e BRAKER1/hintsfile.gff,BRAKER2/hintsfile.gff -o tsebra.gtf
\end{bashinput}

Option \texttt{-g} is used to specify the gene sets that will be combined. The option \texttt{-e} is used to specify the evidence files that will be used for the selection of transcripts. The option \texttt{-o} is used to specify the output file. The output file will contain the combined gene set in \texttt{gtf} format. The additional single-exon gene filter can be manually enabled using the \texttt{-{}-filter\_single\_exon} option.

As a second example, we show how to enforce that all transcripts from one set are kept. This could, e.g., ~be the case if we wish to merge a Galba gene set for a large genome with a BRAKER1 gene set. The BRAKER1 gene set will likely contain a large number of fragmented transcripts due to the genome size, while the Galba gene set has already been filtered with DIAMOND and is likely very reliable.

\begin{bashinput}
singularity exec -B ${PWD}:${PWD} braker3.sif tsebra.py \
  -k Galba/galba.gtf -g BRAKER1/braker1.gtf \
  -e BRAKER1/hintsfile.gff,Galba/hintsfile.gff -o tsebra.gtf
\end{bashinput}

The native output of TSEBRA is in GTF format. It is possible to extract the coding and protein sequences in FASTA format from such a GTF file and the genome file with the following script, which is part of AUGUSTUS:

\begin{bashinput}
singularity exec -B ${PWD}:${PWD} braker3.sif getAnnoFastaFromJoingenes.py \
  -g genome.fa -o tsebra -f tsebra.gtf
\end{bashinput}

This will produce the output files \texttt{tsebra.codingseq} and \texttt{tsebra.aa} in FASTA format.

If you need a GFF3 file, this can also be converted with the following script that is part of AUGUSTUS:

\begin{bashinput}
singularity exec -B ${PWD}:${PWD} braker3.sif cat \ 
  tsebra.gtf | gtf2gff.pl -gff3 --out=tsebra.gff3
\end{bashinput}

\section{Notes}\label{notes}

Both BRAKER and Galba are continuously evolving pipelines. This chapter of the book describes the usage of \texttt{braker.pl} version 3.0.7 and \texttt{galba.pl} version 1.0.11. We recommend using the most recent software versions, as both are actively developed with frequent releases. Check the GitHub repository for the latest versions. Both pipelines have extensive documentation on GitHub. BRAKER is available at \url{https://github.com/Gaius-Augustus/BRAKER} while Galba is available at \url{https://github.com/Gaius-Augustus/GALBA}. TSEBRA is available at \url{https://github.com/Gaius-Augustus/TSEBRA}.

In case of problems, we recommend first checking the extensive list of closed and open issues in the GitHub repositories, as many common problems and questions have already been addressed. If you cannot find a solution to your problem, we recommend opening a new issue in the relevant GitHub repository. The developers of BRAKER and Galba, as well as an active user community, will help you to solve your problem.

We recommend conducting extensive quality control for every final set of genes intended for further research. Visualization of gene structures alongside evidence can be crucial to identify systematic problems in gene prediction. We recommend using the software \texttt{MakeHub} \cite{hoff2019makehub} to generate an Assembly Hub to browse with the UCSC Genome Browser \cite{nassar2023ucsc}. Any other state-of-the-art genome browser will also serve the purpose, but it may require more work to prepare the evidence tracks.

BUSCO is commonly used to describe the completeness of a predicted protein set with respect to clade-specific marker genes. It is important to note that both BRAKER and Galba predict alternative transcript isoforms for such genes. This is often reflected in an increased BUSCO-duplicate rate. To mitigate this problem, select one transcript isoform per gene before running BUSCO. Beware that if evoked with a \texttt{-{}-busco-lineage}, BRAKER will try to minimize the number of missing BUSCOs. This makes the final BUSCO assessment somewhat less valuable for extrapolating the completeness of all genes from the clade-specific marker genes.

OMArk is an interesting addition to the quality check of a gene set. This tool assesses the gene set with respect to a larger number of marker genes and provides additional information, e.g.,~the typical absence of certain genes in the taxonomic lineage. In particular, OMArk supports the correct handling of alternative transcript isoforms.

\section*{Acknowledgement}

This work was built on our previous publication \textit{Whole Genome Annotation with BRAKER} \cite{hoff2019whole}. Both BRAKER and Galba are the result of collaborative international software development. We thank Mark Borodovsky, Alexandre Lomsadze, Mario Stanke, Matthis Ebel, and a larger GitHub user and developer community for their contributions to BRAKER. In particular, we express our gratitude to Neng Huang, who implemented the protein-mode of compleasm for usage with BRAKER, a feature of BRAKER that was not described prior to this chapter. The authors thank Heng Li, Joseph Guhlin, Matthis Ebel, Natalia Nenasheva, Daniel Honsel, Steffen Herbold, Ethan Tolman, Paul Frandsen, and Mario Stanke for their contributions to Galba.

\section{References}
\renewcommand{\refname}{} 
\vspace*{-3em}
\bibliographystyle{unsrt}
\bibliography{literatur}



\newpage

\appendix

\section{Running pipelines with toy data sets}

To descrease runtime, we skip optimizing AUGUSTUS parameters on the example of the toy data sets (resulting parameters are useless for real data because the training data is too small). Since the parameters cannot be reused for any good purpose, we will not specify the \texttt{-{}-species} option.

Wall clock runtime was measured on a HPC system where each node has 72 Intel(R) Xeon(R) Gold 6240 CPU @ 2.60GHz processors. The runtime is given in hours and minutes.

\subsection{Toy example: Running BRAKER3 with a protein database and a BAM file  \protect\includegraphics[width=0.8cm]{figs/braker3.png}\label{braker3bam}}

\begin{bashinput}
wd=toy_braker3 # specify output directory name
T=8 # specify number of threads
singularity exec -B ${PWD}:${PWD} braker3.sif braker.pl --genome=/opt/BRAKER/example/genome.fa \
  --prot_seq=/opt/BRAKER/example/proteins.fa --bam=/opt/BRAKER/example/RNAseq.bam \
  --workingdir=${wd} --threads=${T} --busco_lineage=eukaryota_odb10 --skipOptimize
\end{bashinput}

We report a wall clock runtime of 8 minutes.

\subsection{Toy example: Running BRAKER2 with a protein database \protect\includegraphics[width=0.8cm]{figs/braker2.png} \label{braker2toy}}
\begin{bashinput}
wd=toy_braker2 # specify output directory name
T=8 # specify number of threads
singularity exec -B ${PWD}:${PWD} braker3.sif braker.pl --genome=/opt/BRAKER/example/genome.fa \
  --prot_seq=/opt/BRAKER/example/proteins.fa --workingdir=${wd} --threads=${T} \
  --busco_lineage=eukaryota_odb10 --skipOptimize
\end{bashinput}

We report a wall clock runtime of 18 minutes.

\subsection{Toy example: Running Galba with proteins of closely related species \protect\includegraphics[width=0.8cm]{figs/galba.png} \label{galbatoy}}

\begin{bashinput}
wd=toy_galba # specify output directory name
T=8 # specify number of threads
singularity exec -B ${PWD}:${PWD} galba.sif galba.pl --genome=/opt/GALBA/example/genome.fa \
  --prot_seq=/opt/GALBA/example/proteins.fa --workingdir=${wd} --threads=${T} --skipOptimize
\end{bashinput}

We report a wall clock runtime of 2 minutes.

\subsection{Toy example: Running BRAKER1 with a BAM file \protect\includegraphics[width=0.8cm]{figs/braker1_optional.png} \protect\includegraphics[width=0.8cm]{figs/braker1.png} \label{braker1toy}}

\begin{bashinput}
wd=toy_braker1bam # specify output directory name
T=8 # specify number of threads
singularity exec -B ${PWD}:${PWD} braker3.sif braker.pl --genome=/opt/BRAKER/example/genome.fa \
  --bam=/opt/BRAKER/example/RNAseq.bam --workingdir=${wd} --threads=${T} \
  --busco_lineage=eukaryota_odb10 --skipOptimize
\end{bashinput}

We report a wall clock runtime of 8 minutes.

\subsection{Toy example: Running BRAKER1 with a hints file \protect\includegraphics[width=0.8cm]{figs/braker1.png} \label{braker1toyhints}}

\begin{bashinput}
wd=toy_braker1hints # specify output directory name
T=8 # specify number of threads
singularity exec -B ${PWD}:${PWD} braker3.sif braker.pl --genome=/opt/BRAKER/example/genome.fa \
  --hints=/opt/BRAKER/example/RNAseq.hints --workingdir=${wd} --threads=${T} \
  --busco_lineage=eukaryota_odb10 --skipOptimize
\end{bashinput}

We report a wall clock runtime of 8 minutes.

\section{Running pipelines with fly data}

Full genome scale trained AUGUSTUS parameters may potentially be reused for gene prediction, and we will therefore provide the \texttt{-{}-species} option in the following examples. For simplicity, we will use the same species name as working directory name. Remember that the provided data is not optimized to maximize accuracy. The provided data is only intended to demonstrate the usage of the pipelines.

Runtime was measured on a HPC system where each node has 72 Intel(R) Xeon(R) Gold 6240 CPU @ 2.60GHz processors. The runtime is given in hours and minutes.

\subsection{Fly example: Running BRAKER3 with a protein database and SRA IDs \protect\includegraphics[width=0.8cm]{figs/braker3_optional.png} \protect\includegraphics[width=0.8cm]{figs/braker3.png}\label{braker3sra}}
\begin{bashinput}
T=72 # specify number of threads
wd=fly_braker3_sra # specify output directory name
# retrieve data if not already present
if [ ! -f genome.fa ]; then
  wget https://bioinf.uni-greifswald.de/bioinf/braker/data/genome.fa.gz
  gunzip genome.fa.gz
fi
if [ ! -f Arthropoda.fa ]; then
  wget https://bioinf.uni-greifswald.de/bioinf/partitioned_odb11/Arthropoda.fa.gz
  gunzip Arthropoda.fa.gz
fi
# run BRAKER3
singularity exec -B ${PWD}:${PWD} braker3.sif braker.pl --species=${wd} --genome=genome.fa \
  --prot_seq=Arthropoda.fa --rnaseq_sets_ids=SRR19416937 --workingdir=${wd}$ --threads=${T} \
  --busco_lineage=arthropoda_odb10
\end{bashinput}

The wall clock time is approximately 4 hours and 10 minutes.

\subsection{Fly example: Running BRAKER3 with a protein database and FASTQ files \protect\includegraphics[width=0.8cm]{figs/braker3_optional.png} \protect\includegraphics[width=0.8cm]{figs/braker3.png}\label{braker3fastq}}

\begin{bashinput}
T=72 # specify number of threads
wd=fly_braker3_fastq # specify output directory name
# retrieve data if not already present
if [ ! -f genome.fa ]; then
  wget https://bioinf.uni-greifswald.de/bioinf/braker/data/genome.fa.gz
  gunzip genome.fa.gz
fi
if [ ! -f Arthropoda.fa ]; then
  wget https://bioinf.uni-greifswald.de/bioinf/partitioned_odb11/Arthropoda.fa.gz
  gunzip Arthropoda.fa.gz
fi
if [ ! -d fastq ]; then
  mkdir fastq
fi
cd fastq
if [ ! -f file1_1.fq ]; then
  wget https://bioinf.uni-greifswald.de/bioinf/braker/data/file1_1.fq.gz
  gunzip file1_1.fq.gz
fi
if [ ! -f file1_2.fq ]; then
  wget https://bioinf.uni-greifswald.de/bioinf/braker/data/file1_2.fq.gz
  gunzip file1_2.fq.gz
fi
# run BRAKER3
singularity exec -B ${PWD}:${PWD} braker3.sif braker.pl --species=${wd} --genome=genome.fa \
  --prot_seq=Arthropoda.fa --rnaseq_sets_ids=file1 --rnaseq_sets_dir=${PWD}/fastq/ --workingdir=${wd} \
  --threads=${T} --busco_lineage=arthropoda_odb10
\end{bashinput}

The wall clock time is approximately 4 hour and 20 minutes.

\subsection{Fly example: Running BRAKER3 with a protein database and BAM file \protect\includegraphics[width=0.8cm]{figs/braker3.png}\label{braker3bamfly}}

\begin{bashinput}
T=72 # specify number of threads
wd=fly_braker3_bam # specify output directory name
# retrieve data if not already present
if [ ! -f genome.fa ]; then
  wget https://bioinf.uni-greifswald.de/bioinf/braker/data/genome.fa.gz
  gunzip genome.fa.gz
fi
if [ ! -f Arthropoda.fa ]; then
  wget https://bioinf.uni-greifswald.de/bioinf/partitioned_odb11/Arthropoda.fa.gz
  gunzip Arthropoda.fa.gz
fi
if [ ! -d fastq ]; then
  mkdir fastq
fi
cd fastq
if [ ! -f rnaseq.bam ]; then
  wget https://bioinf.uni-greifswald.de/bioinf/braker/data/rnaseq.bam
fi
singularity exec -B ${PWD}:${PWD} braker3.sif braker.pl --species=${wd} --genome=genome.fa \
	--prot_seq=Arthropoda.fa --rnaseq_sets_dir=${PWD}/fastq/ --rnaseq_sets_ids=rnaseq \ 
	--workingdir=${wd} --threads=${T} --busco_lineage=arthropoda_odb10
\end{bashinput}

The wall clock time is approximately 4 hours and 17 minutes.

\subsection{Fly example: Running BRAKER3 with a protein database and Iso-Seq data \protect\includegraphics[width=0.8cm]{figs/braker3_external.png}\protect\includegraphics[width=0.8cm]{figs/braker3.png} \label{braker3isoseqfly}}

\begin{bashinput}
T=72 # specify number of threads
wd=fly_braker3_isoseq # specify output directory name
# retrieve data if not already present
if [ ! -f genome.fa ]; then
  wget https://bioinf.uni-greifswald.de/bioinf/braker/data/genome.fa.gz
  gunzip genome.fa.gz
fi
if [ ! -f Arthropoda.fa ]; then
  wget https://bioinf.uni-greifswald.de/bioinf/partitioned_odb11/Arthropoda.fa.gz
  gunzip Arthropoda.fa.gz
fi
if [ ! -f isoseq.bam ]; then
  wget https://bioinf.uni-greifswald.de/bioinf/braker/data/isoseq.bam
fi
singularity exec -B ${PWD}:${PWD} braker3_isoseq.sif braker.pl --species=${wd} \
  --genome=genome.fa --prot_seq=Arthropoda.fa --bam=isoseq.bam --workingdir=braker3_isoseq \
  --threads=${T} --busco_lineage=arthropoda_odb10
\end{bashinput}

The wall clock time is approximately 2 hours and 34 minutes.

\subsection{Fly example: Running BRAKER2 with a protein database \protect\includegraphics[width=0.8cm]{figs/braker2.png} \label{braker2fly}}

\begin{bashinput}
T=72 # specify number of threads
wd=fly_braker2 # specify output directory name
# retrieve data if not already present
if [ ! -f genome.fa ]; then
  wget https://bioinf.uni-greifswald.de/bioinf/braker/data/genome.fa.gz
  gunzip genome.fa.gz
fi
if [ ! -f Arthropoda.fa ]; then
  wget https://bioinf.uni-greifswald.de/bioinf/partitioned_odb11/Arthropoda.fa.gz
  gunzip Arthropoda.fa.gz
fi
singularity exec -B ${PWD}:${PWD} braker3.sif braker.pl --species=${wd} --genome=genome.fa \
  --prot_seq=Arthropoda.fa --workingdir=${wd} --threads=${T} --busco_lineage=arthropoda_odb10
\end{bashinput}

The wall clock time is approximately 3 hour and 23 minutes.

\subsection{Fly example: Running Galba with proteins of closely related species \protect\includegraphics[width=0.8cm]{figs/galba.png} \label{galbafly}}

\begin{bashinput}
T=72 # specify number of threads
wd=fly_galba # specify output directory name
# retrieve data if not already present
if [ ! -f genome.fa ]; then
  wget https://bioinf.uni-greifswald.de/bioinf/braker/data/genome.fa.gz
  gunzip genome.fa.gz
fi
if [ ! -f proteins.fa ]; then
  wget https://bioinf.uni-greifswald.de/bioinf/partitioned_odb11/proteins.fa.gz
  gunzip proteins.fa.gz
fi
singularity exec -B ${PWD}:${PWD} galba.sif galba.pl --species=${wd} --genome=genome.fa \
  --prot_seq=proteins.fa --workingdir=${wd} --threads=${T}
\end{bashinput}

We report a wall clock runtime of 7 hours and 46 minutes.

\subsection{Fly example: Running BRAKER1 with a BAM file \protect\includegraphics[width=0.8cm]{figs/braker1_optional.png} \protect\includegraphics[width=0.8cm]{figs/braker1.png} \label{braker1fly}}

\begin{bashinput}
T=72 # specify number of threads
wd=fly_braker1_bam # specify output directory name
# retrieve data if not already present
if [ ! -f genome.fa ]; then
  wget https://bioinf.uni-greifswald.de/bioinf/braker/data/genome.fa.gz
  gunzip genome.fa.gz
fi
if [ ! -f rnaseq.bam ]; then
  wget https://bioinf.uni-greifswald.de/bioinf/braker/data/rnaseq.bam
fi
singularity exec -B ${PWD}:${PWD} braker3.sif braker.pl --species=${wd} --genome=genome.fa \
  --bam=rnaseq.bam --workingdir=${wd} --threads=${T} --busco_lineage=arthropoda_odb10
\end{bashinput}

The wall clock time is approximately 1 hour and 30 minutes.

\subsection{Fly example: Running BRAKER1 with a hints file \protect\includegraphics[width=0.8cm]{figs/braker1.png} \label{braker1hintsfly}}

\begin{bashinput}
T=72 # specify number of threads
wd=fly_braker1_hints # specify output directory name
# retrieve data if not already present
if [ ! -f genome.fa ]; then
  wget https://bioinf.uni-greifswald.de/bioinf/braker/data/genome.fa.gz
  gunzip genome.fa.gz
fi
if [ ! -f hints.gff ]; then
  wget https://bioinf.uni-greifswald.de/bioinf/braker/data/hints.gff
fi
singularity exec -B ${PWD}:${PWD} braker3.sif braker.pl --species=${wd} --genome=genome.fa \
  --hints=hints.gff --workingdir=${wd} --threads=${T} --busco_lineage=arthropoda_odb10
\end{bashinput}

The wall clock time for the fly data set is approximately 1 hour and 43 minutes.

\end{document}